\documentclass{article}

\usepackage[nonatbib,final]{neurips_data_2022}
\usepackage{graphicx}
\usepackage{array}
\usepackage{amsmath}
\usepackage{enumitem}
\usepackage{caption}
\usepackage{subcaption}
\usepackage{booktabs}

\usepackage{todonotes}
\usepackage{color}
\definecolor{darkgreen}{rgb}{0,0.5,0}
\definecolor{purple}{rgb}{1,0,1}
\usepackage{hyperref}[backref=page]
\hypersetup{
    colorlinks=true,
    linkcolor=blue,
    filecolor=magenta,      
    urlcolor=cyan,
}
\newcommand{\websitelink}{\href{https://www.nathanlct.com/research/nocturne}{nathanlct.com/research/nocturne}}

\usepackage[utf8]{inputenc} % allow utf-8 input
\usepackage[T1]{fontenc}    % use 8-bit T1 fonts
\usepackage{url}            % simple URL typesetting
\usepackage{booktabs}       % professional-quality tables
\usepackage{amsfonts}       % blackboard math symbols
\usepackage{nicefrac}       % compact symbols for 1/2, etc.
\usepackage{microtype}      % microtypography
\usepackage{xcolor}         % colors

\title{Nocturne: a scalable driving benchmark for bringing multi-agent learning one step closer to the real world}

\author{%
  Eugene Vinitsky\thanks{\textbf{These authors contributed equally to this work.}} \thanks{Corresponding Author} \\
  Meta AI, UC Berkeley\\
  \texttt{vinitsky.eugene@gmail.com} \\
   \And
  Nathan Lichtlé$^{*}$\\
  UC Berkeley \\ École des Ponts ParisTech\\
  \texttt{nathan.lichtle@gmail.com} \\
     \And
  Xiaomeng Yang$^{*}$\\
  Meta AI\\
  \texttt{yangxm@fb.com} \\
     \And
  Brandon Amos\\
  Meta AI\\
  \texttt{bda@fb.com} \\
     \And
  Jakob Foerster\\
  Univeristy of Oxford\\
  \texttt{jakob.foerster@eng.ox.ac.uk} \\
}

\begin{document}

\maketitle

\begin{abstract}%   <- trailing '%' for backward compatibility of .sty file
We introduce \textit{Nocturne}, a new 2D driving simulator for investigating multi-agent coordination under partial observability. The focus of Nocturne is to enable research into inference and theory of mind in real-world multi-agent settings without the computational overhead of computer vision and feature extraction from images. Agents in this simulator only observe an obstructed view of the scene, mimicking human visual sensing constraints. Unlike existing benchmarks that are bottlenecked by rendering human-like observations directly using a camera input, Nocturne uses efficient intersection methods to compute a vectorized set of visible features in a C++ back-end, allowing the simulator to run at $2000+$ steps-per-second. Using open-source trajectory and map data, we construct a simulator to load and replay arbitrary trajectories and scenes from real-world driving data. Using this environment, we benchmark reinforcement-learning and imitation-learning agents and demonstrate that the agents are quite far from human-level coordination ability and deviate significantly from the expert trajectories. Code for Nocturne is available at \href{https://github.com/facebookresearch/nocturne}{https://github.com/facebookresearch/nocturne}.
\end{abstract}

\section{Introduction}

\begin{figure}
     \centering
     \begin{subfigure}[b]{0.4\textwidth}
         \centering
         \includegraphics[width=0.95\textwidth]{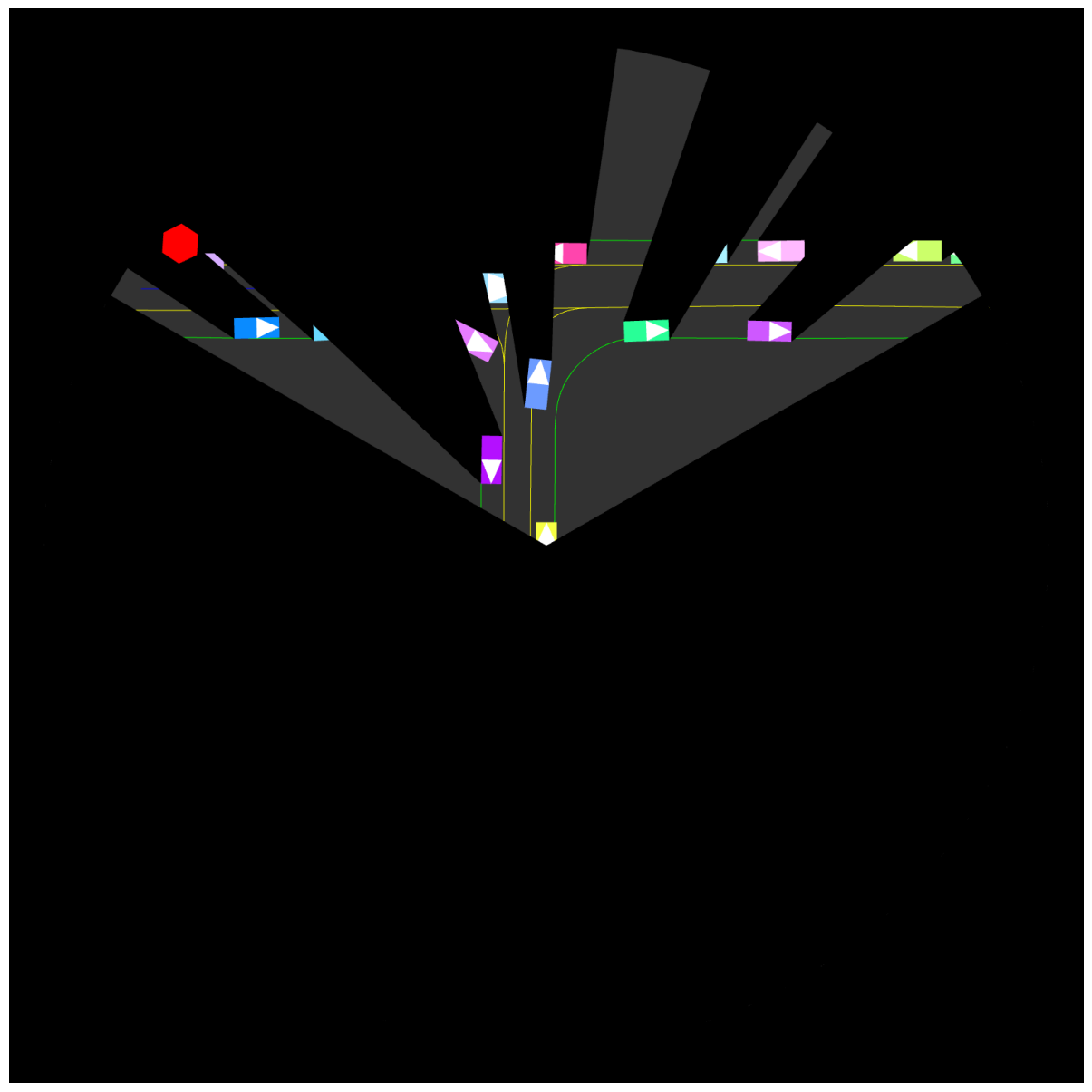}
     \end{subfigure}
    %  \begin{subfigure}[b]{0.32\textwidth}
    %      \centering
    %      \includegraphics[width=0.95\textwidth]{figures/scenes/tfrecord-00358-of-01000_60_sourceid_21_features.png}
    %  \end{subfigure}
     \begin{subfigure}[b]{0.4\textwidth}
         \centering
         \includegraphics[width=0.95\textwidth]{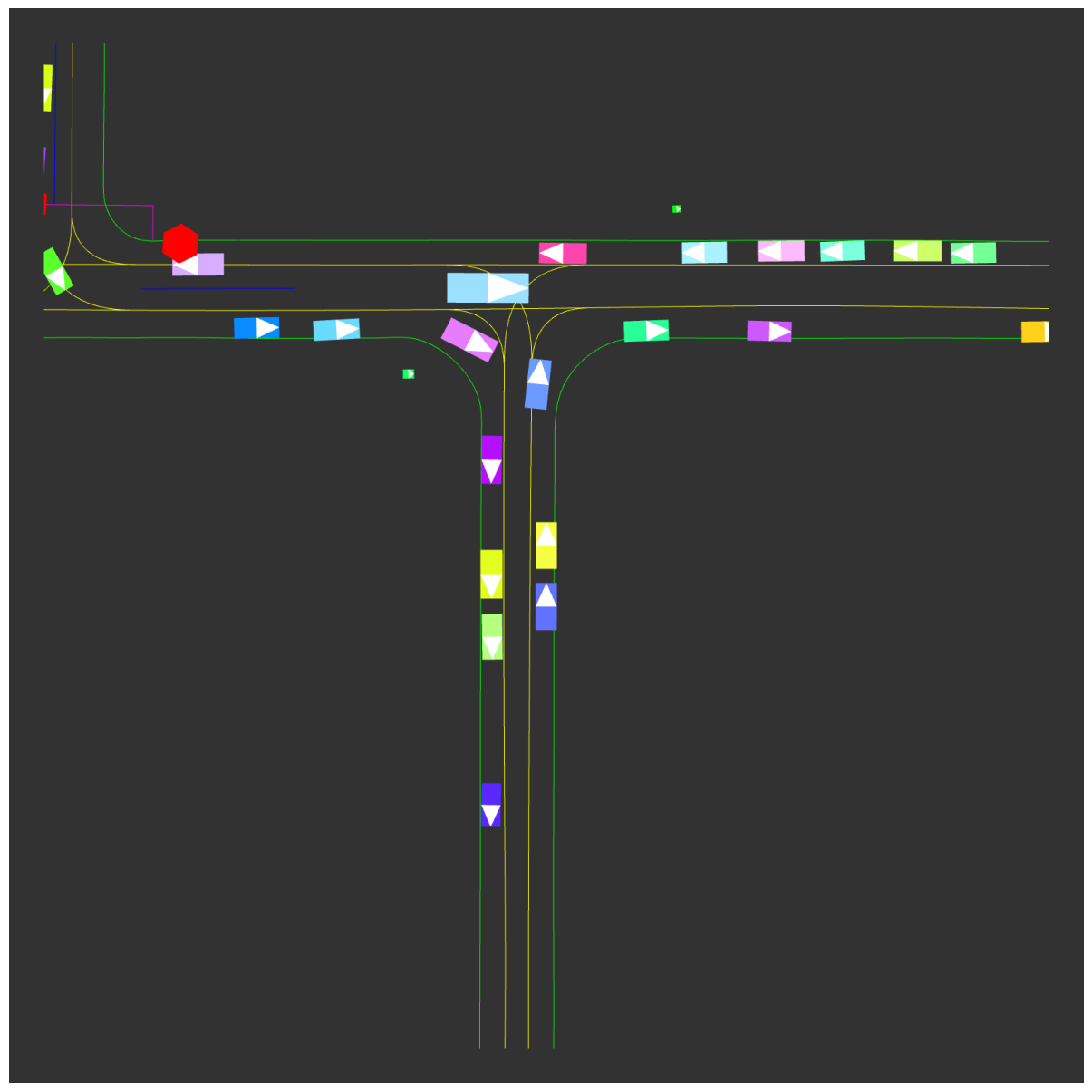}
     \end{subfigure}
    \caption{A visual depiction of the obstruction model used to represent the objects that are visible to the agents. 
    %(Left) Unobstructed view of the original scene centered on the yellow agent. 
    %(Middle) Obstructed view with the viewing yellow agent in the center of the cone. 
    % (Right) Visual representation of which features the yellow agent would get in the obstructed case.
    (Left) Obstructed view with the viewing yellow agent in the center of the cone. (Right) Original scene centered on the cone agent with an unobstructed view.}
    \label{fig:agent_observability}
\end{figure}

This paper presents \textit{Nocturne}, a new simulator and benchmark for multi-agent driving under human-like sensor uncertainty that is intended to aid the process of studying real-world multi-agent coordination and learning. Instead of combining the challenges of coordination with feature extraction from images, Nocturne is a 2D simulator that generates vector representations of the set of objects and road points that would be visible to an idealized human driver (see Fig.~\ref{fig:agent_observability} for an example) and supports head-tilt to acquire additional information about blind spots. In contrast to driving benchmarks that achieve partial-observability by using a camera input, Nocturne uses efficient visibility-checking methods and a C++ back-end to enable us to construct observations and step the dynamics of a single agent at thousands of steps-per-second (see Appendix Sec.~\ref{app:sec:speed} for an exact analysis). This speed is key to its use in multi-agent learning settings where frequently billions of environment interactions are needed to learn performant agents~\cite{bard2020hanabi, leibo2021scalable}. In contrast to many existing multi-agent learning benchmarks, Nocturne is neither zero-sum nor fully-cooperative but mixed-motive, combining the challenges of coordination and cooperation.

Crucially, Nocturne is not just a simulator. Instead, it is built upon open-source driving data and features a diverse set of real-world scenes (see Fig.~\ref{fig:diverse_scenes}) that probe the ability of agents to safely navigate and coordinate in complex scenes such as intersections, roundabouts, parking lots, and highways.
We use this data as a source of experts for imitation, to flexibly vary the number of controlled agents in the scene, and as a train-test split for validating the generalization ability of a human-driver model.

The challenge we propose is to learn (or otherwise design) policies that achieve the same set of final states as human experts (throughout this work we will call this the \textit{goal rate} and the final state the \textit{goal position}) while achieving a 0\% collision rate. Achieving a high goal rate alongside a negligible collision rate is challenging for any policy design scheme (both learning and non-learning) due to the combination of the high-dimensional state space, the partial observability of the scene, the large number of interacting agents, and the decentralization of the policies at test time. The secondary challenge in Nocturne is to find policies that closely mimic human behavior at the trajectory level. Building on top of the human data we provide an evaluation scheme and off-the-shelf baseline implementations including evaluations.

We provide a guide to the construction and rules of the benchmark as well as results on using Nocturne to design agents and test their capabilities and human-similarity. We demonstrate that learning effective agents in Nocturne is challenging; tuned RL and imitation learning baselines struggle to successfully complete the highly interactive scenes. Finally, we demonstrate that the agents achieve relatively low distance to the expert trajectories and show in Appendix Sec.~\ref{sec:app_zsc} that there does not appear to be a zero-shot coordination problem~\cite{hu2020other} at this level of agent capability.

{\setlength\intextsep{0pt}
\setlength{\belowcaptionskip}{0pt}
\begin{figure}
     \centering
     \begin{subfigure}{0.4\textwidth}
         \centering
         \includegraphics[width=0.85\textwidth]{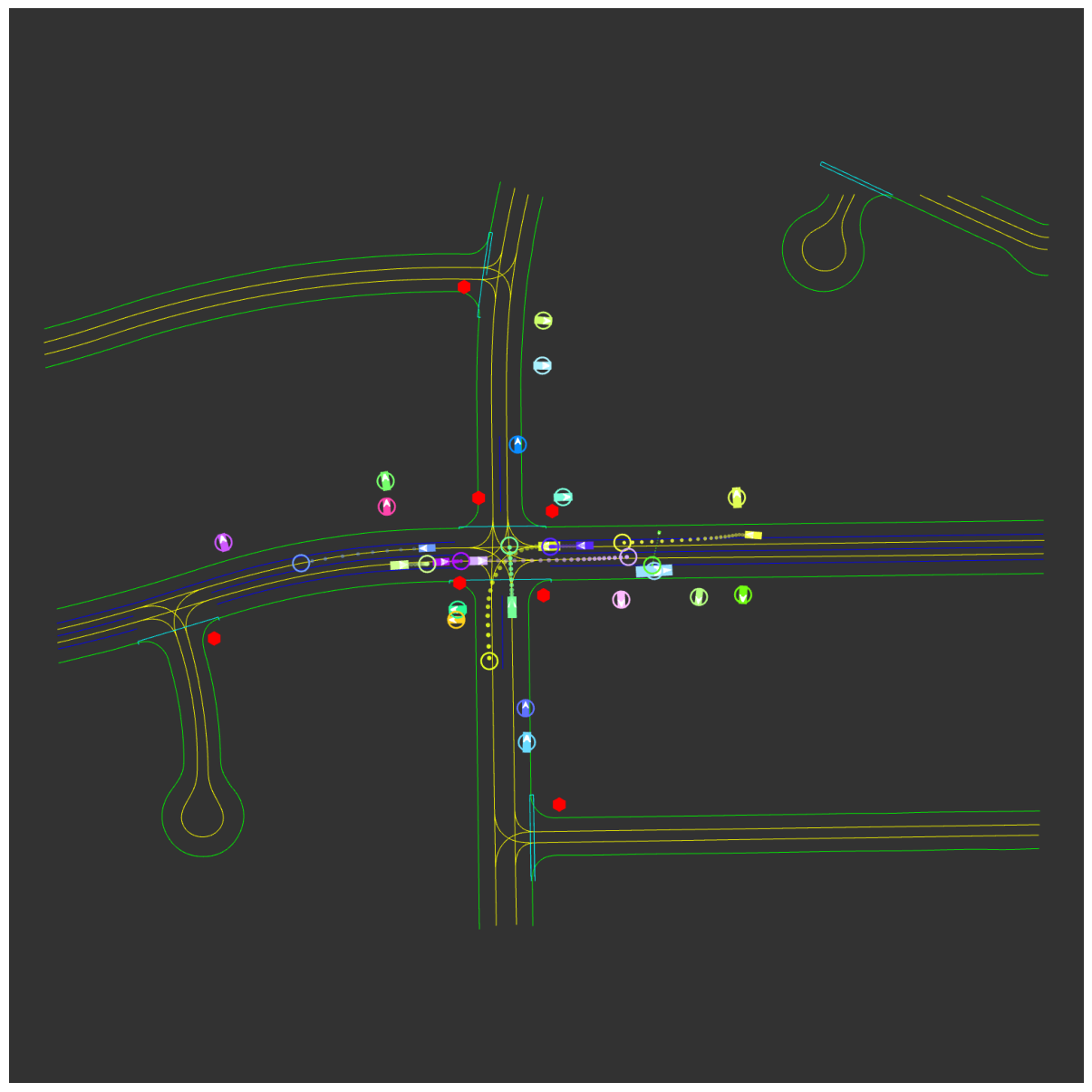}
         \caption{A four-way stop.}
         \label{fig:four_way_stop}
     \end{subfigure}
     \begin{subfigure}{0.4\textwidth}
         \centering
         \includegraphics[width=0.85\textwidth]{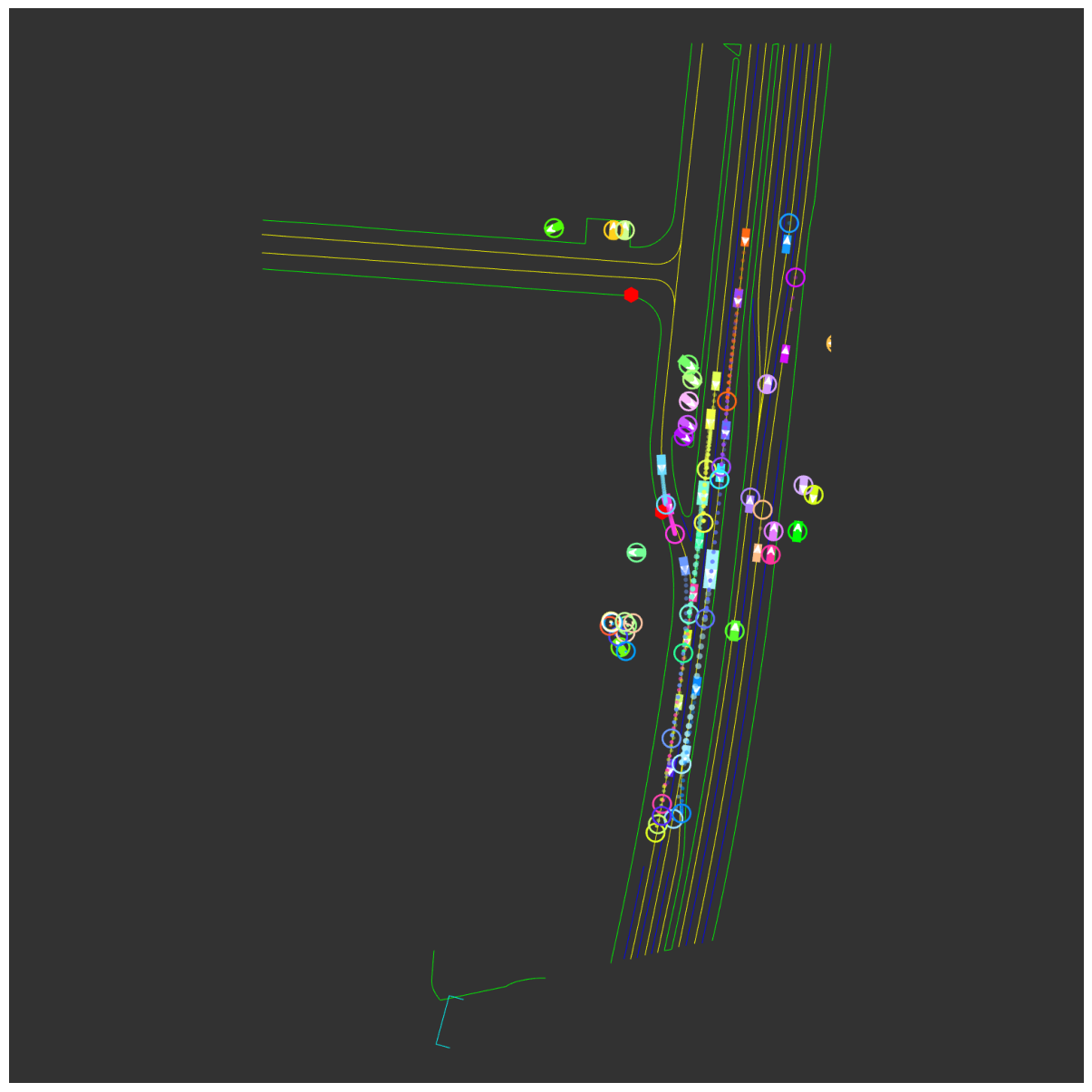}
         \caption{A straight road with merging vehicles.}
         \label{fig:straight_road}
     \end{subfigure}
    \vskip\baselineskip
     \begin{subfigure}{0.4\textwidth}
         \centering
         \includegraphics[width=0.85\textwidth]{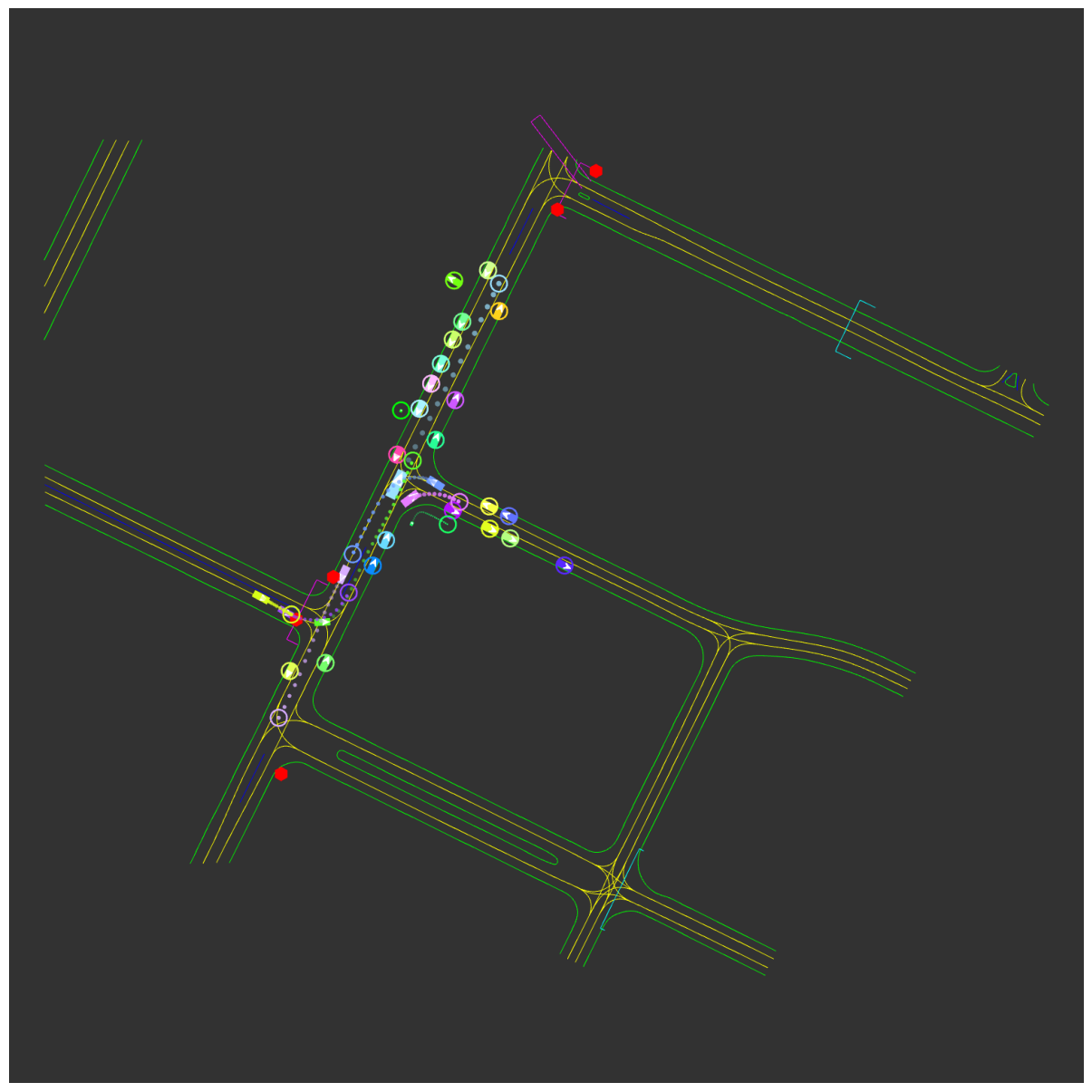}
         \caption{Unprotected turns with conflicting routes.}
         \label{fig:unprotected_turn}
     \end{subfigure}
     \begin{subfigure}{0.4\textwidth}
         \centering
         \includegraphics[width=0.85\textwidth]{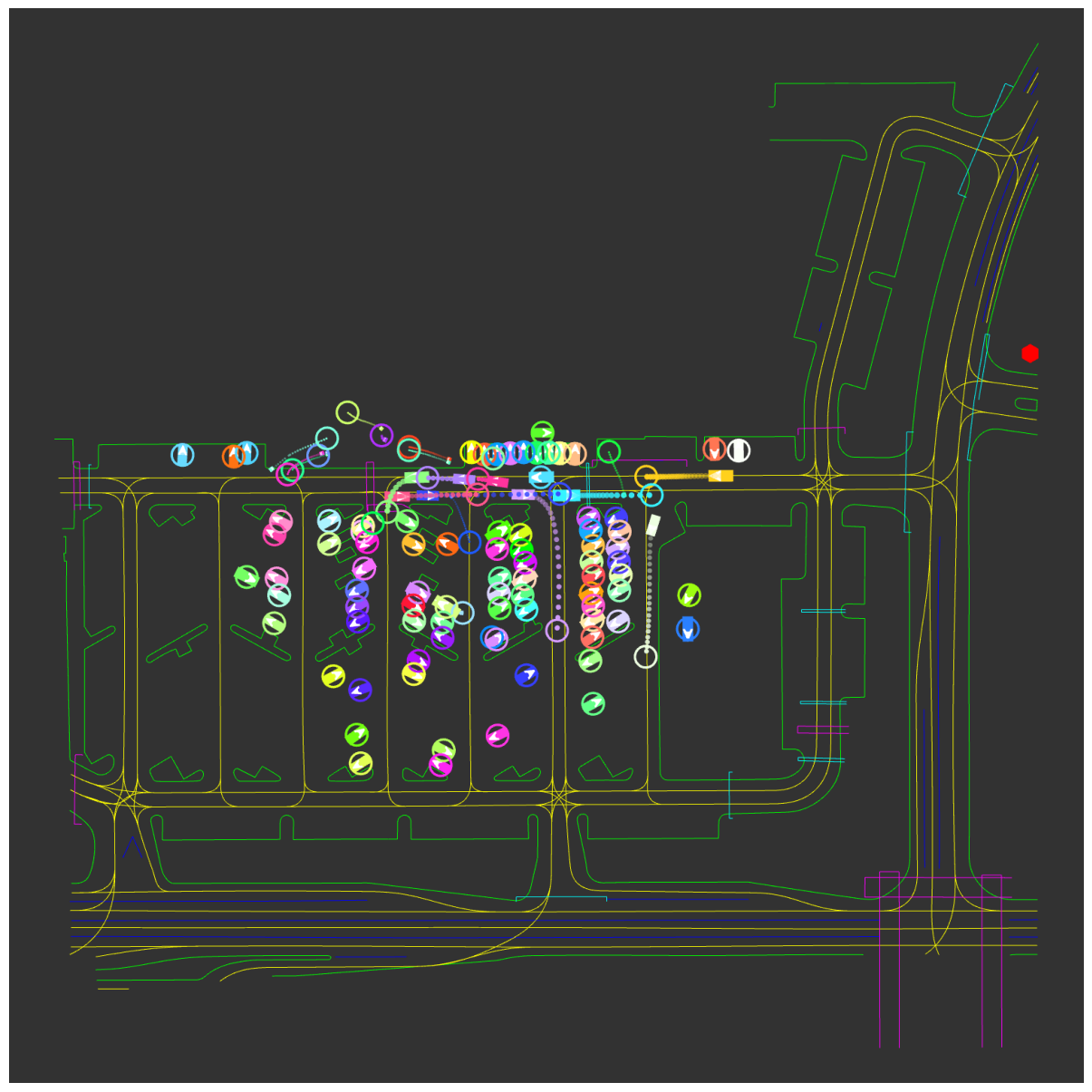}
         \caption{A crowded parking lot.}
         \label{fig:crowded_lot}
     \end{subfigure}
    \caption{Four scenes demonstrating the diversity of the navigable scenes in Nocturne. Colored circles represent the goal position of the corresponding colored agent. Dots represent the trajectory of the agent, with opacity increasing as time goes on. Videos of experts negotiating these scenes can be found at \websitelink.}
    % \vspace{-20mm}%Put here to reduce too much white space after your table 
    \label{fig:diverse_scenes}
\end{figure}
}

\section{Related Work}
\noindent
\textbf{Multi-agent traffic simulation tools and benchmarks:} \\
In terms of multi-agent driving benchmarks that require both acceleration and steering control, there are several closely related benchmarks that differ in terms of ease of implementing multi-agent interaction, being data-driven, 2D vs. 3D, the mechanism by which they support partial observability, or the rate at which they can construct the partially observed state for an agent. We summarize some of these differentiating features in Table~\ref{tab:differentiation}. 

The closest works to ours are BARK~\cite{bernhard2020bark}, SMARTS~\cite{zhou2020smarts}, and MetaDrive~\cite{li2021metadrive}.
SMARTS~\cite{zhou2020smarts} supports multi-agent driving in a wide variety of interactive driving scenarios and offers a large set of default human driving behaviors to use for testing autonomous vehicles. BARK~\cite{bernhard2020bark}, like Nocturne, is a 2D goal-driven simulator with support for external datasets and contains a wide variety of large-scale scenarios and multi-agent support. Finally, MetaDrive~\cite{li2021metadrive} also supports real-world data and multi-agent interaction and is able to achieve 300 FPS image rendering by rendering lower fidelity images. The main differentiating feature from these works is the support for acquiring the set of objects that are visible without requiring the rendering of camera images; to our knowledge Nocturne is the only available simulator that can jointly compute an agent’s visible objects and the agent's dynamics step above 2000 steps-per-second (see Appendix Sec.~\ref{app:sec:speed} for exact details). Additional simulators are summarized in Table ~\ref{tab:differentiation}. 

\begin{table*}[!t]
\centering
\caption{
Comparison of representative driving simulators. State-based partial observability refers to whether the set of visible objects can be queried. Expert data refers to whether recorded trajectory data is available for the scenes. Baseline human models refers to whether explicit models of human driving are available to control road objects in the scene.
}

\label{tab:comparison}
% \resizebox{0.95\linewidth}{!}{%
\begin{tabular}{cccccccccc}
% \centering
\hline
Simulator & 
\begin{tabular}[c]{@{}c@{}}Multi-agent\\ Support \end{tabular} & 
\begin{tabular}[c]{@{}c@{}}2D/3D \end{tabular} & % TODO!
\begin{tabular}[c]{@{}c@{}}State-Based \\ Partial \\ Observability \end{tabular} & 
\begin{tabular}[c]{@{}c@{}}Expert \\ Data \end{tabular} &
\begin{tabular}[c]{@{}c@{}} Baseline \\ Human \\ Models\end{tabular} &\\
% \begin{tabular}[c]{@{}c@{}}SPS \end{tabular} &\\
\hline
CARLA~\cite{dosovitskiy2017carla}        &  &3D&  && \checkmark 
\\ \hline % TODO
SUMMIT~\cite{cai2020summit}        &\checkmark  &3D&  && \checkmark  
\\ \hline % TODO
MACAD~\cite{palanisamy2020multi}        &\checkmark  &3D&  && \checkmark  
\\ \hline% ok
Highway-env~\cite{highway-env}       & &2D& & &\checkmark 
\\ \hline % ok
Sim4CV~\cite{muller2018sim4cv}      & &3D& & & 
\\\hline % TODO
Duckietown~\cite{paull2017duckietown}   & &3D&  & & 
\\ \hline %
SMARTS~\cite{zhou2020smarts}   &\checkmark &2D&  & 
&\checkmark \\ \hline %
MADRaS~\cite{santara2021madras}  &\checkmark &2D&  & 
&\checkmark \\ \hline %
DriverGym~\cite{kothari2021drivergym}  & &2D & &\checkmark & \checkmark 
\\ \hline %
DeepDrive-Zero~\cite{craig_quiter_2020_3871907}   &\checkmark &2D& &  &   \\ \hline %
MetaDrive~\cite{li2021metadrive}    &\checkmark &3D& &\checkmark& \checkmark \\ \hline 
VISTA~\cite{amini2021vista}   &\checkmark &3D& &\checkmark& \checkmark  \\ \hline %
\textbf{Nocturne} &\checkmark  &2D &\checkmark& \checkmark  &  \\ \hline %
\label{tab:differentiation}
\end{tabular}
% }
\end{table*}

\noindent
\textbf{Partially observed multi-agent benchmarks:}\\
There are a wide variety of partially-observable multi-agent benchmarks not focused on driving. Within the card-playing domain, Hanabi~\cite{bard2020hanabi} is frequently used to investigate coordination under partial observability and features between 2-5 agents. The Starcraft multi-agent benchmark (SMAC)~\cite{samvelyan2019starcraft}, perhaps the most ubiquitous MARL benchmark, features many agents and high-dimensional observations but does not come with human data and many algorithms now achieve perfect performance in this challenge~\cite{yu2021surprising}. Melting Pot~\cite{leibo2021scalable} features a huge diversity of many-agent mixed-motive challenges but does not have available expert data. Finally, games such as Poker, Stratego, and Diplomacy are frequently studied.

\noindent
\textbf{Open-Source Trajectory Data:}\\
We provide a brief overview of large driving datasets that contain both map information and annotated trajectory data for vehicles, pedestrians, and cyclists. Available trajectory data can be categorized along the size of the dataset, the diversity of data in the dataset, the method of collection, and the available annotations of the scene (e.g. maps, traffic lights, road object trajectories). This work uses the Waymo Motion dataset~\cite{ettinger2021large} as it has a high diversity of scenes across many cities and contains annotations for cyclists, pedestrians, and traffic lights, features that will be used in future versions of Nocturne. However, we note that Argoverse 2~\cite{wilson2021argoverse} and nuScenes~\cite{caesar2020nuscenes} have similar features and could have been used; nuScenes also has complete traffic light annotations instead of annotations only for ego-observed traffic lights. The Lyft Level 5 dataset~\cite{houston2020one} appears to be the largest available dataset and has thorough road and vehicle annotations but is drawn from a single stretch of road. For non-egocentric datasets, the INTERACTION dataset~\cite{zhan2019interaction} contains a diversity of scenes across both cities and countries, thereby capturing a variety of driving styles and norms, but is an order of magnitude smaller than other datasets.

\color{black}
\section{Benchmark construction}
\subsection{Defining a Nocturne Scene}
In the following sections, we will refer to objects that can move (vehicles, pedestrians, cyclists) as \emph{road objects} and anything that cannot move (lane lines, road edges, stop signs, etc.) as \emph{road points}. \emph{Road points} are connected together to form a polyline. We will refer to the type of road polyline that should not be crossed by vehicles as a \emph{road edge}.

Nocturne scenes require a map consisting of polylines, a set of initial and final road object positions, and optionally a set of trajectories for the road objects.
Nocturne currently acquires its scenes, goals, and expert trajectories from the Waymo Motion dataset~\cite{ettinger2021large} but can be easily configured to support any dataset that represents its road features as points or polylines. The Waymo Motion dataset consists of $487004$ nine-second trajectory snippets discretized at a rate of 10 Hz with the first second intended to be used as context and the latter eight seconds to be used for prediction.

One challenge of selecting and constructing the scenarios for Nocturne is that these trajectories are collected by labeling the vehicles observed by a Waymo car as it drives. Consequently, we do not have a complete birds-eye view of the scene.
Hence, there are cars that may have been in the scene that were not visible to the Waymo vehicle and therefore are not included in the dataset; for the same reason, the expert trajectories are also incomplete and may not persist throughout the entire duration of the rollout. In other words, the expert trajectories contain agents that unpredictably flicker in and out of existence.
Similarly, the set of traffic lights that were visible to the Waymo vehicle may be insufficient to uniquely determine the underlying traffic light state. For this reason, this first version of the Nocturne benchmark comes with a few restrictions: we do not use traffic lights and we only include vehicles available at the first time-step of the scene to avoid collisions induced by unpredictable vehicle appearances. We filter out scenes that contain traffic lights which leaves the majority of the remaining scenes as roundabouts, unsignalized intersections, arterial roads, and parking lots. This leaves the benchmark consisting of $134453$ snippets.

% \subsubsection{Road Objects}

% \subsubsection{Road Points}
% TODO describe type of road lines

\subsection{Partial Observability Model and Collision Handling}
\label{sec:partial_obs_rules}
To support the investigation of coordination under partial observability, agents in Nocturne come with a configurable view-cone as is shown in Fig.~\ref{fig:agent_observability}. An element (road object or road point) is considered observable if there is a single ray in the cone that intersects with that element that does not pass through any road object on its path to the element. Stop signs are always visible if they are within the agent view-cone even if they would be obscured on the assumption that they would be raised at a sufficiently high level. We select the angle of the cone to be 120 degrees (approximately the range of binocular vision) and 80 meters radius. We note that this does not perfectly mimic a human visual model which has additional complexities such as dynamic variations in the functional field of view~\cite{crundall1999driving}, phenomena relating to interactions between visible objects such as crowding~\cite{bouma1970interaction}, and the necessary ability to pay attention to objects further than 80 meters away at high speeds. Furthermore, we currently represent our vehicles as rectangles and so this model of partial observability will be overly pessimal and neglect situations such as the ability for a driver to sometimes see over the hood of a car or 3-D effects like tall cars seeing over small ones.

The primary challenge in computing which road points and objects are observable is their relatively large number: a scene contains about $30$ vehicles and $4800$ road edge points on average. We use a Bounding Volume Hierarchy (BVH) to maintain the road objects and select candidates for potentially observed vehicles. We build the BVH using approximate agglomerative clustering~\cite{gu2013efficient} to generate a high-quality BVH. When computing the observed objects, we first use the BVH to select the candidates that lie in the \emph{axis-aligned bounding box} (AABB) of the conic view field. Since there are a comparatively larger number of all types of road points (about $16000$ on average), we use a 2D range tree~\cite{bentley1978decomposable} to maintain all of the road points. When computing the observed road points, we do a range search in the 2D range tree to select the candidates that lie in the AABB of the conic view field. For both vehicles and road points, once we have the candidates in the conic view-field, we perform a brute-force visibility check for the object and road points respectively by ray-casting from the viewing agent to all the candidate vertices. Finally, these data-structures are similarly used to accelerate collision checking. 

% We use a similar procedure to the visibility checking to accelerate our collision detection of vehicle-vehicle and vehicle-road collisions. For vehicle-road collisions, we consider a vehicle to have collided if its body intersects with any line segment of the polylines that constitute the road edges. For both vehicle-vehicle and vehicle-road collision detection, we query the road object BVH and a separate road line segment BVH (formed once at the beginning of the episode) to generate candidates that are then evaluated for collision.

\subsection{Construction of the Partially Observable Stochastic Game}
\subsubsection*{Definition of the state space}
Nocturne supports two possible state representations: a rasterized image and a vectorized representation of that image. While we provide default state representations, we consider it fair game on the benchmark to use any other representation as long as only objects that are visible under the conditions in Sec.~\ref{sec:partial_obs_rules} are presented to the agent. Conforming to these consistent rules about visibility allows for a fair comparison between different algorithmic approaches.

For the default vectorized representation, we adopt a fully descriptive set of features (speed, angle, width, length, etc.) for the road objects and use the VectorNet~\cite{gao2020vectornet} representation for the road points. We will refer to the vehicle whose observation is being returned as the \emph{ego vehicle}.
Note that by default all features that can be placed into relative coordinates are returned in relative coordinates to the ego vehicle (e.g. speed is relative speed, heading is relative heading, etc). The agent goal is set to be the final position, speed, and heading of the expert agent. An agent is considered to have achieved its goal if it is within $1$ meter of the final position and within $1 \,\frac{\text{m}}{\text{s}}$ of the final speed of the agent and $0.3 \,\text{radians}$ of the final heading when the target position tolerance is achieved. For exact observations for the ego object, road objects, and road points, as well as the padding mechanism used to handle variations in the number of observed objects / points / stop signs see Appendix Sec.~\ref{sec:exp_details}.

\subsubsection*{Action Space}
Vehicles are driven by acceleration and steering commands that are passed to a bicycle model to update the vehicle state. In the experiments used in this paper we use 6 discrete actions for acceleration, 21 discrete actions for steering, and 5 discrete actions for head tilt with the acceleration actions uniformly splitting $\left[-3, 2\right] \,  \frac{m}{s^2}$, the steering actions between $\left[-0.7, 0.7 \right] \, \text{radians}$, and the head tilt between $\left[-1.6, 1.6 \right] \, \text{radians}$. For more details on the vehicle model, see Appendix Sec.~\ref{sec:vehicle_model}. % Sec. ~\ref{app:veh_model}.

\subsubsection*{Environment Dynamics and Goals}
The total length of the expert data is $9$ seconds, discretized into steps of size $0.1$ seconds. For the first 1 second of the episode, all vehicles obey the expert policy. This is used to construct a history of observations for each agent that can be used to initialize or warm up the policy. After this transitory period, the episode continues for a fixed length of $T=80$ steps. Agents are provided with a target position and target speed that are taken from the final speed and position of the expert trajectories. If an agent achieves their goal they receive a reward of $T$ and are removed from the system. A vehicle will also be removed if it collides with any road edge or object.

In addition, there is a process for selecting the set of vehicles that are controlled in the environment. First, we only control vehicles that at some point have a speed above $0.05 \frac{\text{m}}{\text{s}}$ and that are more than $0.2 \text{m}$ from their goal; in general, these are vehicles that need to move to get to their goal. From this set of vehicles, we remove all vehicles that are already at their goal. Next, we randomly select up to a maximum of $20$ of them and set the remainder to replay expert trajectories. This latter process for keeping a maximum number of controlled vehicles is used in this paper due to constraints of the RL library and is not an aspect of the benchmark.

\subsection{Rules of the Benchmark}
\label{sec:rules}
We outline here a few rules that we expect solvers to respect to ensure consistency between solutions.
\begin{itemize}
    \item[--] The size of the view cone is fixed to $120$ degrees and a distance of $80$ meters. This ensures consistency between the level of partial-observability each controller must handle. Users can also tilt the head of the driver by up to $90$ degrees in either direction to acquire more information.
    \item[--] Only the first $1$ second of the trajectory can be used as context or to warm-start a memory-based controller; control of the vehicles must start at $1$ second into the trajectory. 
    \item[--] A trajectory that successfully reaches the goal is only considered valid if it reaches the goal within the $8$ second time-window. All the scenes are easily solved without a time-constraint by simply creeping forwards slowly.
    \item[--] The environment comes with default rewards and observations but any amount of reward-shaping or observation sharing at training time is valid. However, at test time only information that is directly observable to the agent can be used as input to the policy. For example, sharing information about other agents' goals would be valid at train time but not at test time.
    % While the default observation built into the benchmark does not include information about the map outside of the agent view, adding additional map infor
    \item[--] Adding additional map information to the agent state space beyond the information provided by default is valid. This corresponds to humans often having knowledge about map layout beyond what is within eyesight.
    \item[--] The bounds on the action space should be respected: the acceleration should be bounded between $[-6, 6] \, \frac{\text{m}}{\text{s}^2}$, the change in heading should not be faster than $40$ degrees per second, and driver head tilt should be maintained within $90$ degrees in either direction. This rules out solutions that rapidly get to their goal by using accelerations that are outside of possible vehicle speed bounds or excessively sharp turns.
\end{itemize}
We note that these rules may make some of the scenes unsolvable. For example, real human drivers are not randomly initialized into a scene with a maximum of 1 second of prior context; this context may be critical to safely navigating the scene under the time constraints. Additionally, our model of human perception is not an exact match for true human perception which can extend well beyond 80 meters under certain circumstances. It is possible that there may be missing context at this viewing distance which is crucial for safe navigation. We view these constraints as similar to the challenge of label noise in supervised datasets; our constraints may place an upper limit on the percentage of goals that can be achieved but the benchmark still constitutes a valid comparison between methods as long as the constraints are respected.

\subsection{Unusual features of the Benchmark}
\label{sec:oddities}
For completeness, we note a few oddities of the benchmark that we believe are critical for potential solution designers to be aware of. These challenges are often properties of labeling noise and the egocentric view under which the data was collected.

\subsubsection*{Filtered out pedestrians, cyclists, and traffic lights}
While the dataset contains scenes with traffic lights, pedestrians, and cyclists, the first set of Nocturne environments, NocturneV1, operates solely on scenes that have been filtered to not include traffic lights. Additionally, when constructing the environment, we remove all pedestrians and cyclists from the scene. There is still an option to include them in the visible state, which we do for behavioral cloning. Future versions of the benchmark will consider joint learning of pedestrians, cyclists, and vehicles but for NocturneV1 we only consider vehicles as appropriate modeling of pedestrian and cyclist dynamics is challenging.

\subsubsection*{Egocentric data collection}
The Waymo Motion dataset that forms the basis of the first version of Nocturne is collected by driving a sensor-equipped car and recording the trajectories of all visible vehicles. Thus, in the original data vehicles may have trajectories that are shorter than the full 9 seconds and may only appear midway through the trajectory or appear and disappear throughout the trajectory when they are obscured from the view of the sensing car. Rather than suddenly teleport cars into the scene midway through an episode, we choose to only use cars that have valid states beginning of the episode. This reduces the total number of vehicles that might appear in the episode but does not change the feasibility of any of the agent goals.

\subsubsection*{Infeasible goals}
Roughly $3\%$ of the vehicles in the dataset have an expert trajectory that crosses an impassible road edge. This is due to labeling errors in the dataset where, for example, small gaps in road edges are occasionally missed that make the road edge crossable. To ensure a benchmark where all goals are achievable, we compute all trajectories where crossing a road edge was necessary to achieve the goal and set these vehicles to replay their expert trajectory rather than be controlled; this corresponds to $3\%$ of all vehicles. Finally, $2\%$  of vehicles in the dataset are initialized in a colliding state. This primarily occurs in parking lot scenes where a vehicle slightly overlaps with the boundary of a parking spot or a nearby vehicle. These vehicles are also removed. For the calculations of these statistics, see Appendix Sec.~\ref{sec:infeasible}.

\subsubsection*{Illogical goals}
Occasionally, agents may have goals that seem illogical; for example, agents may be asked to come to a full stop in the middle of a highway. While these goals may seem odd, they are actual states that were achieved by humans driving on the roadways. These odd goals are often a consequence of unobserved objects such as drivers queued up that make the vehicles stop so as to avoid a collision that was not observed by the egocentric data collection process.

\color{black}
\section{Experiments Setup}
We run a reinforcement learning and an imitation learning baseline to demonstrate that these tasks do not appear to be easily solved even after billions of steps. We also briefly investigate whether the policies appear to have a zero-shot coordination challenge wherein policies perform well when paired with agents from their seed but are incompatible with policies from a different training run.

We test the following methods:
\begin{itemize}
    \item APPO~\cite{petrenko2020sample} trained in multi-agent mode with a shared policy i.e. every agent is controlled by the same policy but in a decentralized fashion.
    % \item Asynchronous proximal policy optimization (APPO)~\cite{petrenko2020sample} trained in single-agent mode i.e. we randomly select one agent that is actuated and replay the rest from expert data. (Experiments done, data not processed yet)
    \item Behavior Cloning~\cite{bain1995framework, schaal1996learning}.
\end{itemize}
For the RL method, in addition to the fixed-bonus for achieving the goal, we add the following dense reward to encourage the agent to make progress toward the target position, speed, and heading:
\begin{equation}
    r_t = 0.2 \times \left(1 - \frac{||x_t - x_g||_2}{||x_0 - x_g||_2}\right) + 0.2 \times \left(1 - \frac{||v_t - v_g||_2}{40}\right) + 0.2 \times \left(1 - \frac{f(h_t, h_g)}{2 \pi}\right)
\end{equation}
where $x_t$ is the position at time-step $t$, $x_g$ is the goal position, $v_t$ is the speed at time-step $t$ and $v_g$ is the target speed, $h_t$ and $h_g$ are the current and target heading in world coordinates (i.e. not in a relative frame). $f(h_t, h_g)$ returns the minimum angle between $h_t$ and $h_g$. Since coordinates are in a relative frame and the agent cannot observe the world frame, note that $f(h_t, h_g)$ is an observation provided to the agent. The use of this dense reward, which certainly affects the form of the optimal policy, is not a necessary component of the benchmark and here is just used to generate good policies.

All experiments are evaluated on the same $200$ randomly selected scenes from the validation set and statistics are computed over an $8$ second rollout. The width in all the plots represents twice the standard error. For more details and hyperparameters for both methods and architectures, see Appendix Sec.~\ref{sec:exp_details}.

\section{Results and Analysis}
\subsection{Success rate of baselines}
\label{sec:analysis_baseline}
\begin{figure}
    \centering
         \begin{subfigure}[b]{0.32\textwidth}
         \centering
             \includegraphics[width=\textwidth]{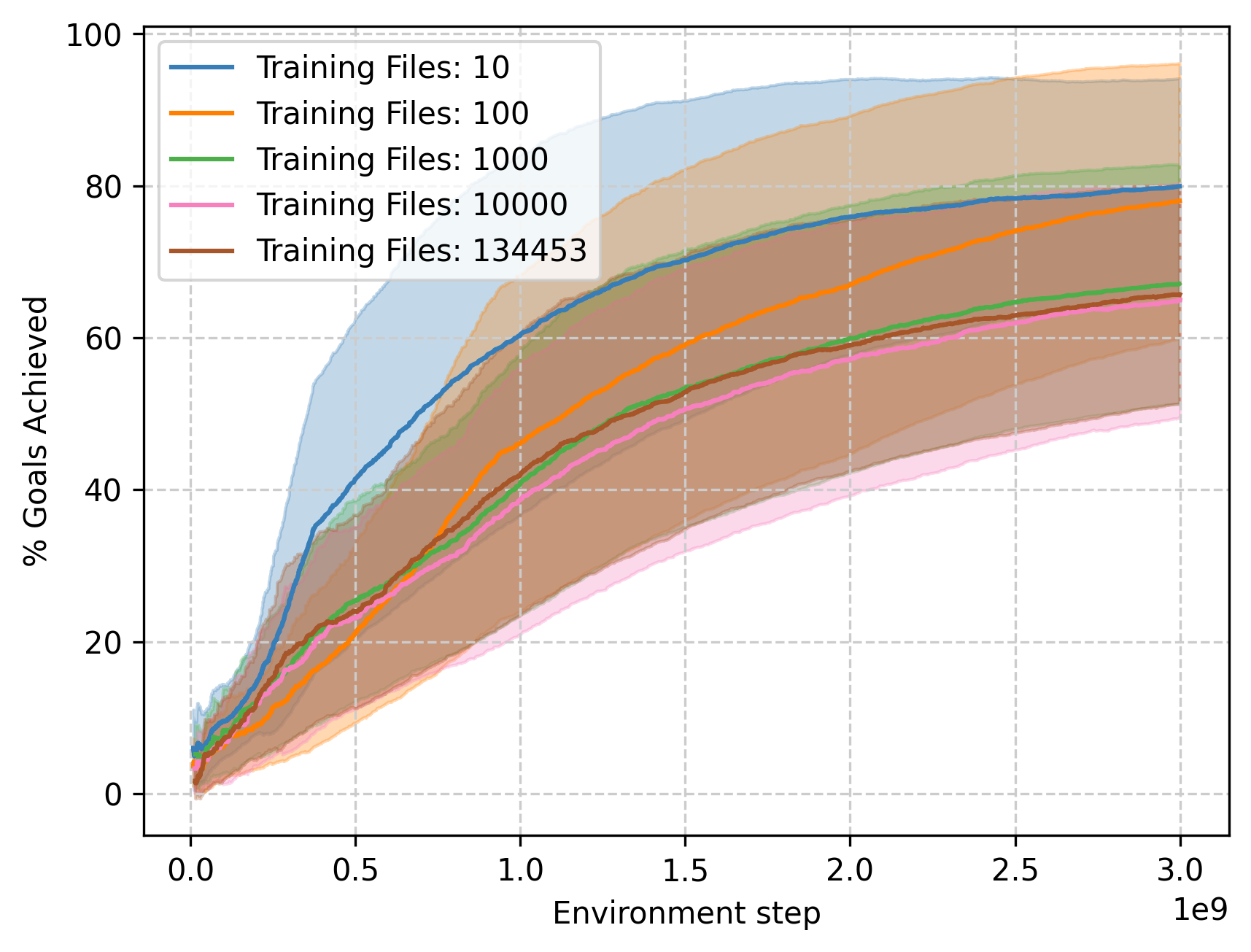}
     \end{subfigure}
     \begin{subfigure}[b]{0.32\textwidth}
         \centering
         \includegraphics[width=\textwidth]{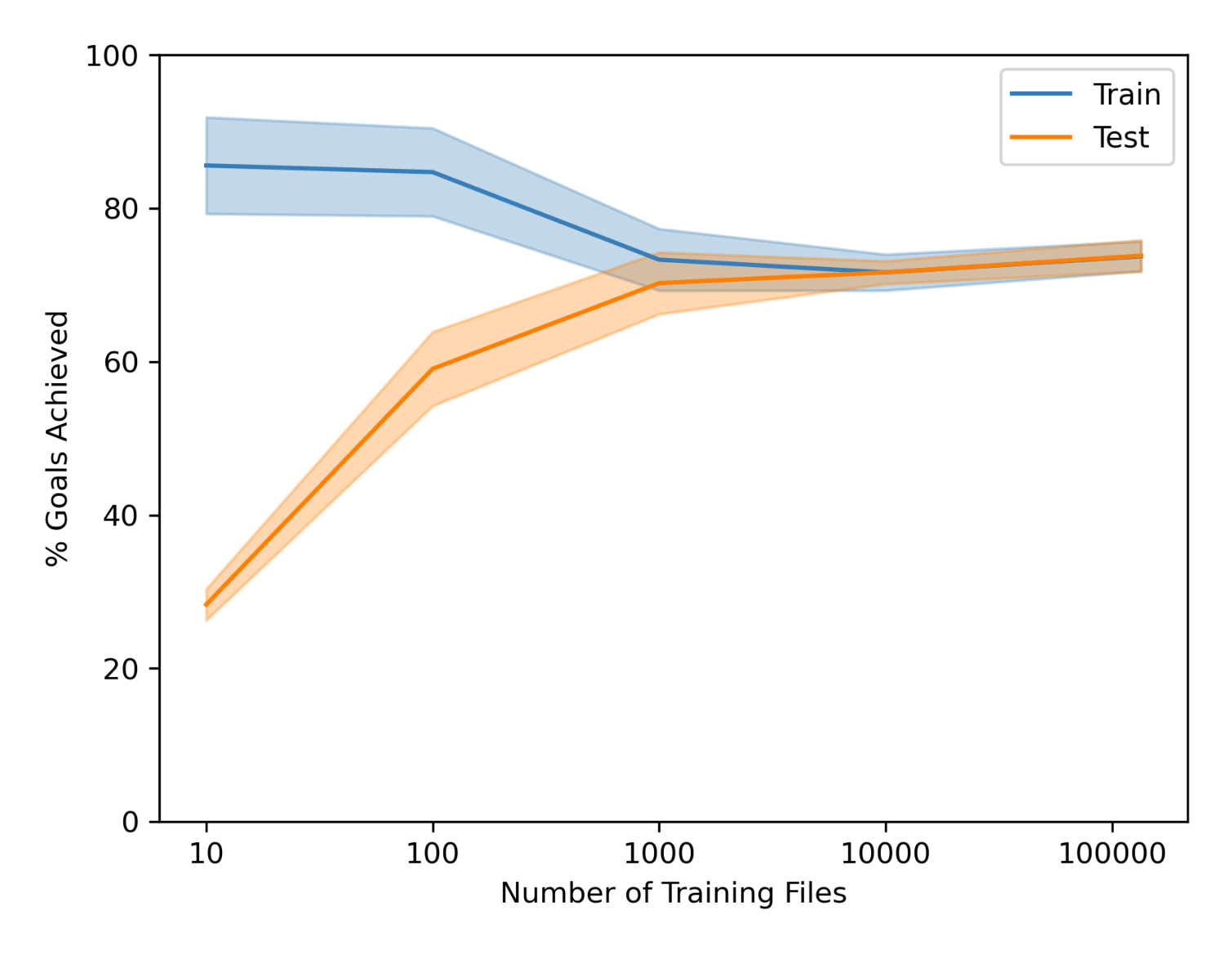}
     \end{subfigure}
     \begin{subfigure}[b]{0.32\textwidth}
         \centering
         \includegraphics[width=\textwidth]{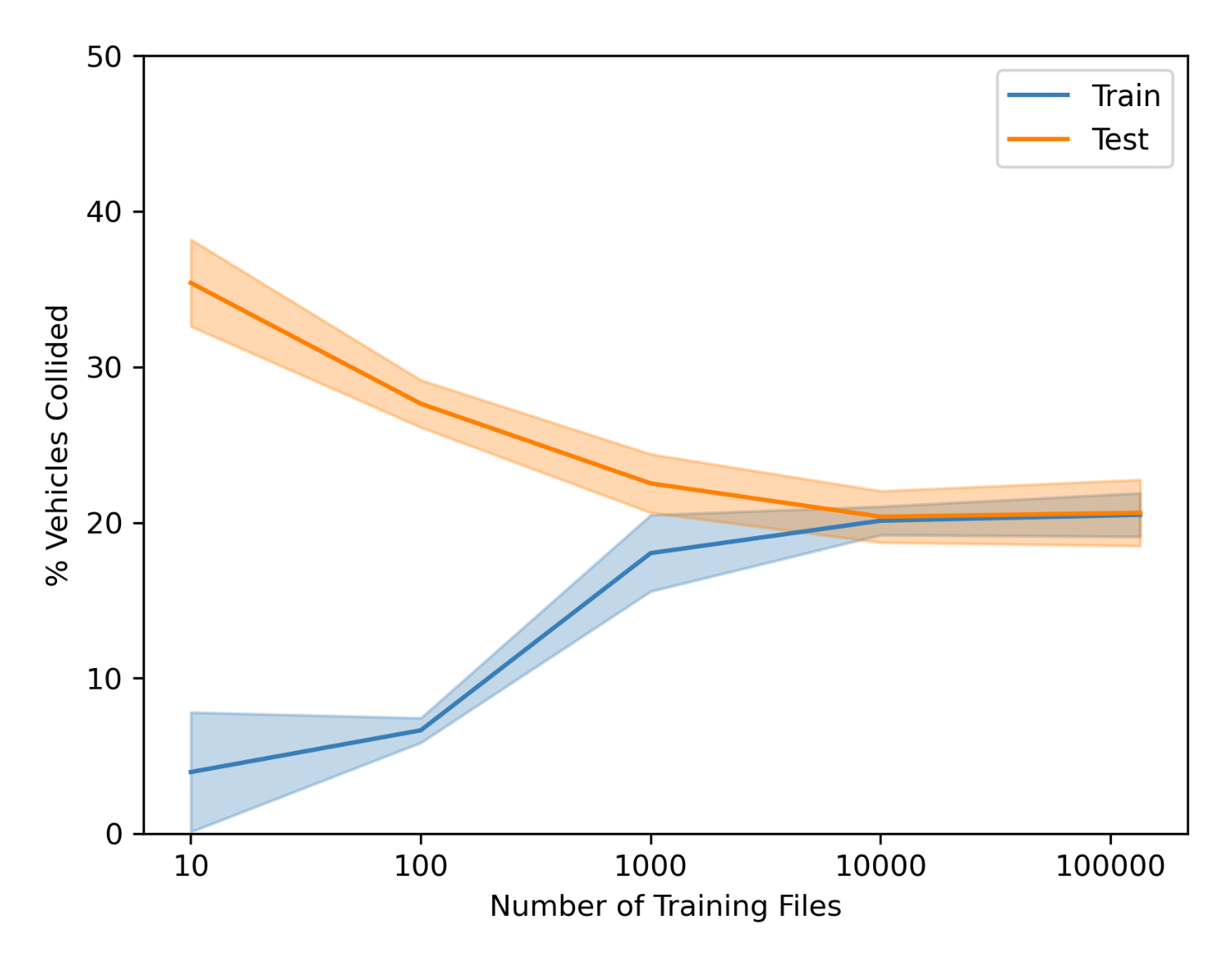}
     \end{subfigure}
    \caption{(Left) Success at getting to the specified goal on the training data as a function of number of environment steps. "Training Files: X" means the agent was trained on X fixed scenes sampled from the training dataset. (Middle) Percent of agents that achieved their goals. (Right) Percent of agents that collided.}
    \label{fig:goal_err}
\end{figure}

Fig.~\ref{fig:goal_err} shows the performance of the agents on the training set after $3e9$ steps which takes approximately two days on 1 GPU and 10 CPUs. Each line corresponds to a policy trained on a fixed subset of the training scenes. We score our policies in two primary ways: the fraction of vehicles that achieve their goal (goal rate) and the fraction of vehicles that collide with another vehicle or road edge (collision rate). Note that this use of collision rate differs from its use in papers such as~\cite{igl2022symphony, suo2021trafficsim} which compute collision rate as the fraction of scenes that contain a vehicle-vehicle collision.

We investigate the effect of the size of the dataset on the train and test performance. In the right half of Fig.~\ref{fig:goal_err} we can see that the inclusion of a larger training dataset decreases the performance of the algorithms up to $1000$ files as they struggle with the diversity of the data. However, the inclusion of additional data narrows the magnitude of the train and test gap and closes it fully at $10000$ files. However, it is still possible that a divergence between train and test might re-emerge as agents become more capable on the train set and approach a $100\%$ goal rate.

Finally, Table~\ref{table:comparison} compares the APPO and BC agents. For the APPO agent, we include the results for the agent trained on $10000$ training files. The expert playback row refers to replay of the expert trajectories.

\begin{table}[ht]
\caption{Overview of metrics across methods for an $8$ second rollout.} % title of Table
\centering % used for centering table
\begin{tabular}{c c c c c} % centered columns (4 columns)
\hline\hline %inserts double horizontal lines
Algorithm\rule{0pt}{1ex} & Collision Rate (\%) & Goal Rate (\%) & ADE (m) & FDE (m) \\  % inserts table
%heading
\hline % inserts single horizontal line
Expert Playback & $4.9$ & 100 & 0 & 0 \\ % inserting body of the table
APPO & $20.3 \pm 0.8$ & $71.7 \pm 0.7$ & $3.1 \pm 0.2$ & $6.1 \pm 0.3$ \\
BC & $38.2 \pm .1$ & $25.3 \pm 0.1$ & $5.6 \pm 0.1$ & $9.2 \pm 0.1$\\
\hline %inserts single line
\end{tabular}
\label{table:comparison} % is used to refer this table in the text
\end{table}

\subsection{Human-agent trajectory similarity}
\begin{figure}
    \centering
     \begin{subfigure}[b]{0.37\textwidth}
         \centering
         \includegraphics[width=\textwidth]{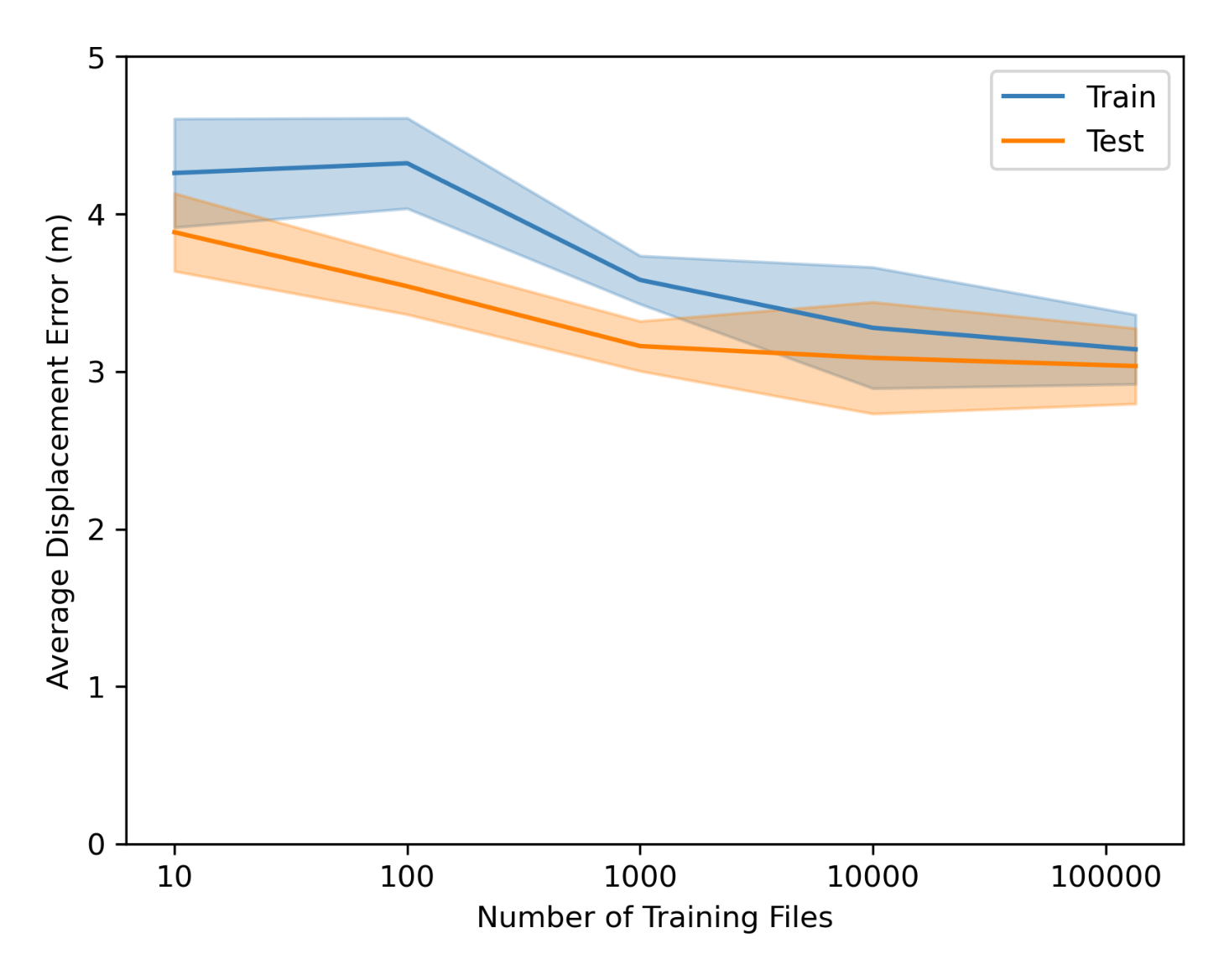}
     \end{subfigure}
          \begin{subfigure}[b]{0.37\textwidth}
         \centering
         \includegraphics[width=\textwidth]{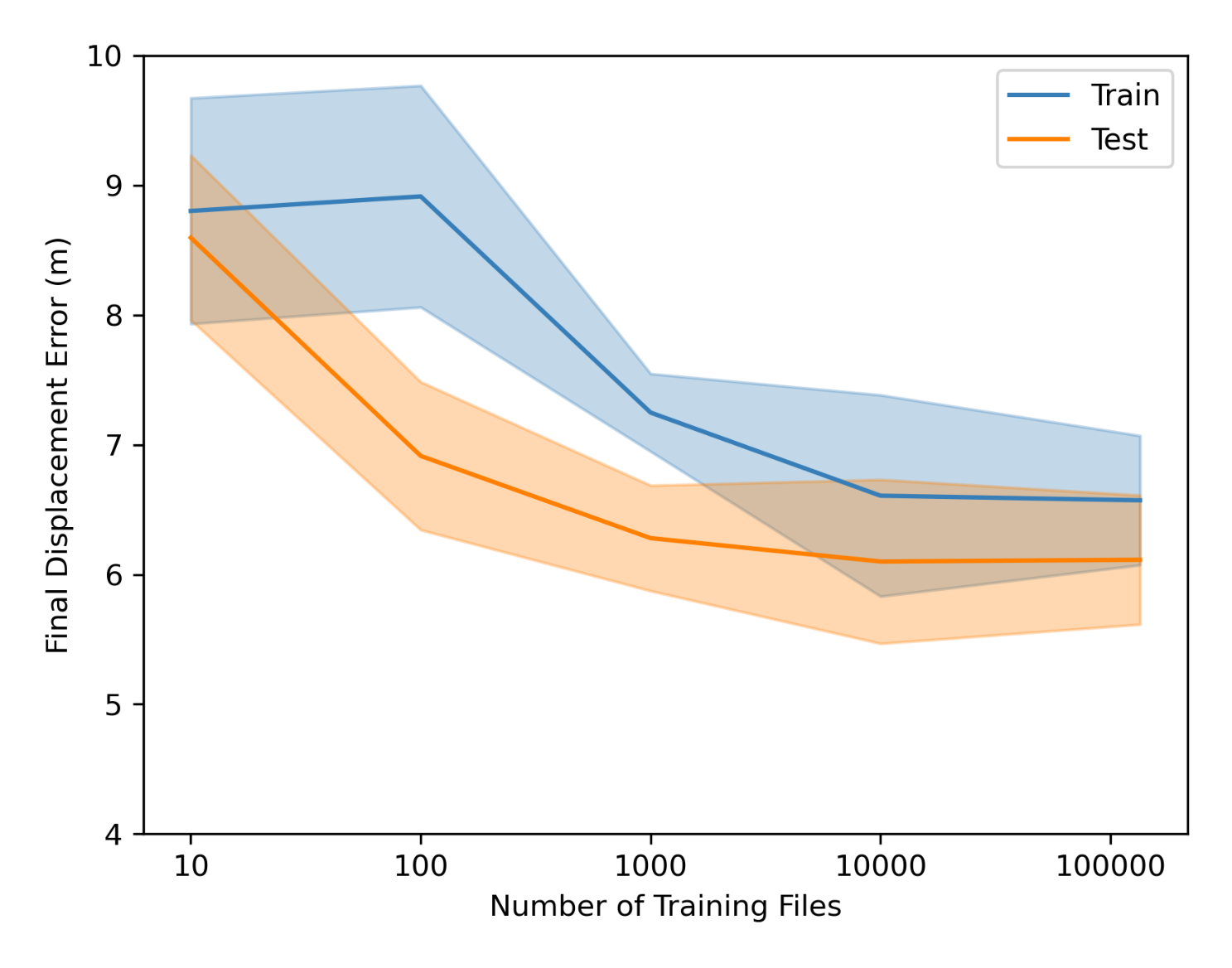}
     \end{subfigure}
    \caption{(Left) Average displacement error (mean l2-distance between an agent and an expert at each time-step). (Right) Final displacement error (l2-distance between an agent and an expert at the final time-step that an expert has a valid state).}
    \label{fig:ade_plot}
\end{figure}

We analyze the results of our experiments with respect to how human-like the resultant policies are using displacement error between expert and agent trajectories as our metric (i.e. L2 distance between the agent and expert trajectories). To align with the definition used in other works~\cite{igl2022symphony, suo2021trafficsim} we disable the removal of vehicles upon collision / reaching the goal. However, we note a few dissimilarities that make comparisons with other works difficult. First, agents are provided with a goal position, a feature that is often not available to other predictive methods. Second, our experts are stepped in scenes that may contain pedestrians or cyclists but our agents are replayed in the same scene without the corresponding pedestrians or cyclists. This can make the magnitude of the displacement error not directly comparable to the values in other works; the low value of the displacement error may simply indicate that a majority of the scenes have unique optima. 
Fig.~\ref{fig:ade_plot} examines the average difference in position between the agent and expert trajectories averaged across 5 training runs and demonstrates that the influence of more training data on displacement error is flat after 1000 files.

% Finally, Fig.~\ref{fig:zsc} compares the collision rate when we randomly sample half the agents in the scene from one policy and half from another. We observe a sharp coordination problem, the agents from one training run are not compatible with the agents from another training run.
% \begin{figure}
%     \centering
%     \includegraphics[width=0.8\textwidth]{temp_figures/zsc_cross_play.png}
%     \caption{Caption}
%     \label{fig:zsc}
% \end{figure}

\subsection{Policy Failure Modes}
\label{sec:failure_modes}
\begin{figure}
    \centering
     \begin{subfigure}[b]{0.37\textwidth}
         \centering
         \includegraphics[width=\textwidth]{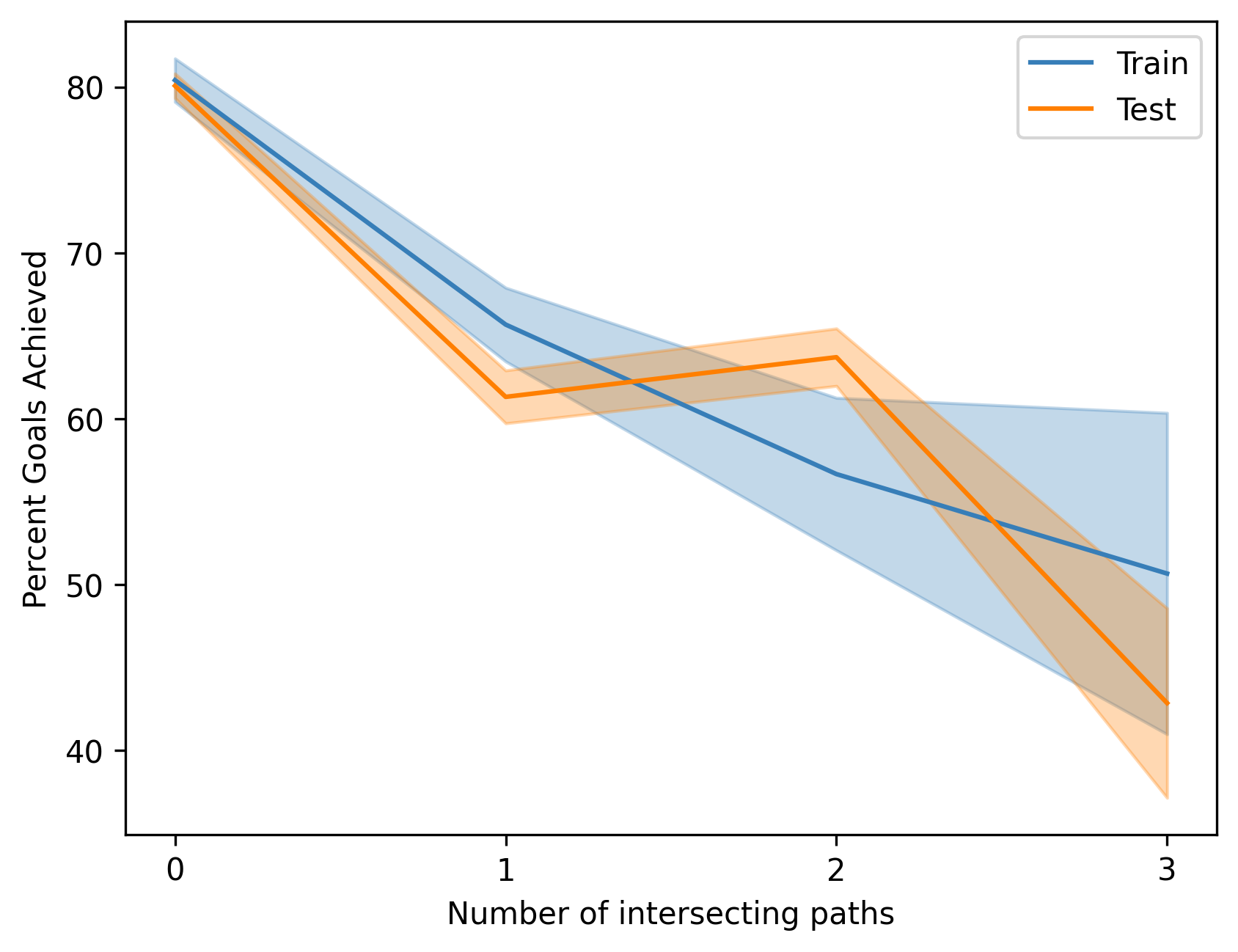}
     \end{subfigure}
          \begin{subfigure}[b]{0.37\textwidth}
         \centering
         \includegraphics[width=\textwidth]{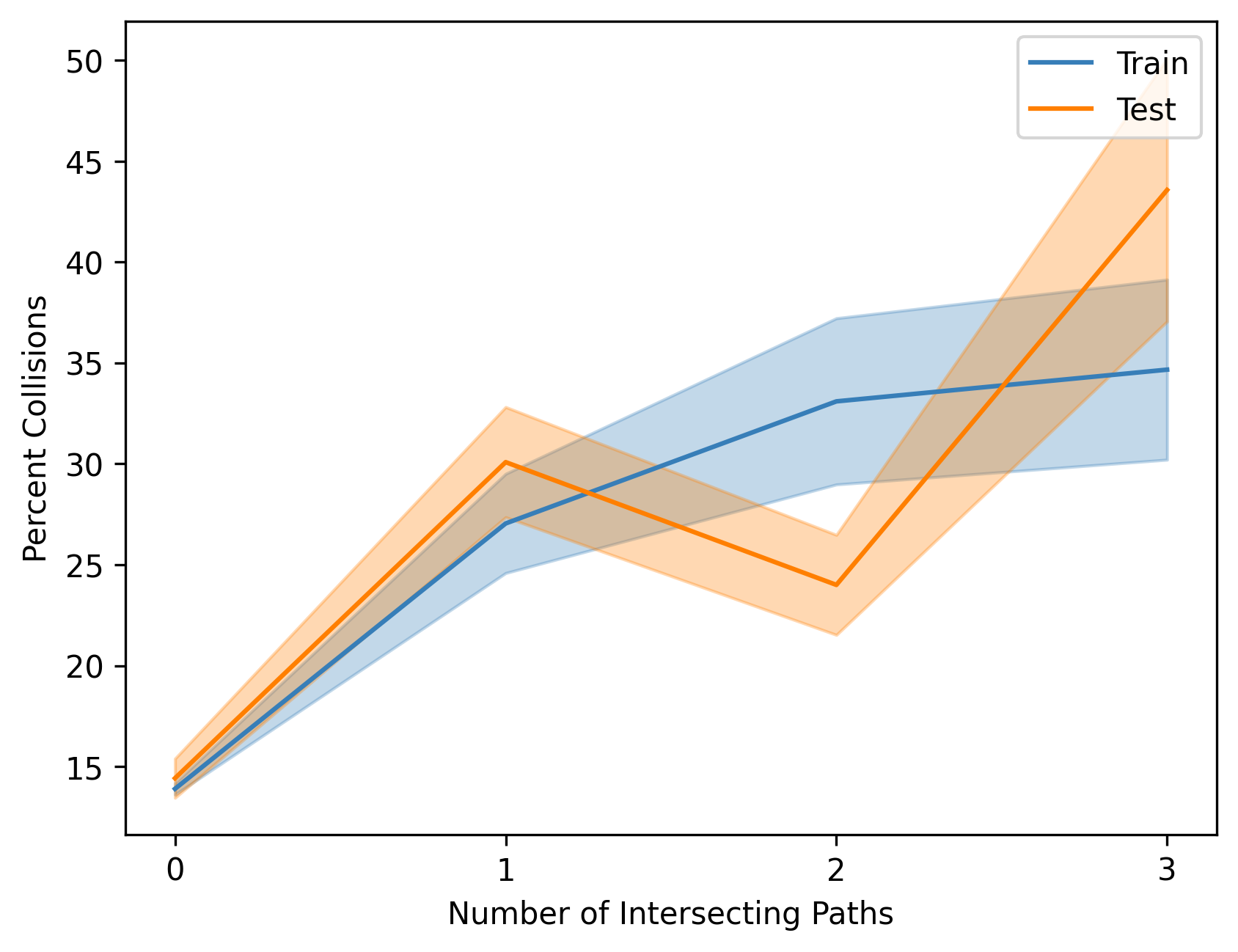}
     \end{subfigure}
    \caption{Goal rate (left) and collision rate (right) of vehicles as a function of the number of times that their corresponding expert trajectory intersected with another expert trajectory (intersections). As more than 3 interactions are rare, creating noisy statistics, scenes with more than 3 interactions are placed into the 3 interaction bin.}
    \label{fig:interactions}
\end{figure}

Here we investigate the mechanisms under which our policies fail to achieve their goals and collide to shine a light on potential avenues for improvement. The key failure modes we qualitatively observe are failures in scenes in which agents are required to interact with another agent either by waiting or merging. Videos of some of the failure modes can be seen at \websitelink. While we cannot measure interactivity directly, we measure a proxy by looking at the intersection of the expert trajectories. We play the experts forwards, record their trajectory as polylines, and consider a vehicle to have as many interactions as there are intersections of the vehicle's expert trajectory polyline with other vehicle's expert trajectory polylines (i.e. if two vehicles' trajectories in time cross at a point, that's an interaction). This captures interactions such as crossing at a four-way stop, merges, and others but neglects interactions such as complementary left turns and also may unintentionally pick up behaviors such as driving behind another vehicle. Close to $25\%$ of vehicles have at least one interaction. The collision and goal rates as a function of interactions are plotted in Fig.~\ref{fig:interactions} and demonstrate that the goal rate declines precipitously and the collision rate increases sharply as the number of interactions increases. This suggests that our agents have learned to get to their goals but perform poorly in settings where getting to the goal requires coordination with another agent.

\section{Conclusion}
We have introduced Nocturne, a simulator and benchmark intended to aid in the study of human-like decentralized coordination for driving systems. We present results on the applications of RL and imitation to this system, however, there still remains work to be done to build agents that operate with the collision and goal rate that humans achieve as well as how to learn these agents efficiently. Given better agents, human-like rules and conventions may be emergent properties of driving safely in these settings~\cite{pal2020emergent}.

There is also ample remaining work to be done on new benchmarks. Due to the ego-centric data collection, we are forced to remove vehicles once they achieve their goal (i.e. the last observed position of the driver in the data). However, it may be possible to use generative models or other generative mechanisms to sample new goals for the agents to continue their trajectory once they achieve the goals set out in the data. Similarly, generative models could be used to complete the traffic light states and enable the inclusion of the traffic light scenes.

Finally, one open question is how to use Nocturne agents as predictive models of human driving. At the moment a Nocturne agent requires a goal to which it is driving. To use Nocturne agents for prediction (say for an autonomous vehicle trying to predict the motion of agents in the scene), a method for inferring goals from the 1-second context needs to be implemented. A topic for future work is to use supervised methods to predict the goals and enable fully decentralized prediction of Nocturne agents from egocentric observations.

\section{Reproducibility and Ethical Statement}
We do not release trained models as the Waymo Motion dataset restricts the release of trained models. While the dataset contains images that have been anonymized, only trajectory data is used in this work. The benchmark and all files needed to run it are publicly available.

\section{Acknowledgements}
We thank the Python community~\cite{van1995python} for creating the core tools that enabled our
work including Hydra~\cite{yadan2019hydra}, Pytorch~\cite{paszke2019pytorch}, Matplotlib~\cite{hunter2007matplotlib}, and numpy~\cite{oliphant2006guide, van2011numpy}. A huge thanks to Aleksei Petrenko for a ton of support in tuning and debugging SampleFactory~\cite{petrenko2020sample}. Thanks to Scott Ettinger for help in understanding some of the peculiarities of the Waymo Motion dataset~\cite{ettinger2021large}. Thanks to Rachit Singh for help with some of the results analysis scripts. We would also like to thank the International Emerging Actions project SHYSTRA (CNRS) for support of Nathan Lichtlé.


% \section*{References}
\bibliographystyle{acm}
\bibliography{sample}

\clearpage

\newpage

\appendix
 
\section{Experiment Details}
\label{sec:exp_details}
\subsection{Architecture and computational resources}
The APPO experiments are each run for two days on 1 V100 GPU and 10 CPUs (Intel Xeon CPU E5-2698 v4 @ 2.20GHz) and each experiment is run for 5 seeds. For the RL experiments we run 5 conditions over 5 seeds for 2 days leading to a total of $1200$ GPU hours and $12000$ CPU hours. For the Imitation Learning experiments we run $5$ seeds for three hours leading to $15$ GPU hours and $100$ CPU hours.

The architecture for the APPO agent is a two-layer neural network with 256 hidden units and Tanh non-linearities followed by a Gated Recurrent Unit~\cite{cho2014learning} with 256 hidden states. The Behavior Cloning experiment is run on a single GPU for five seeds, trained on 1000 files, for 600 gradient steps with a batch size of 512. Here instead of using a recurrent neural network we stack the current state and the prior 4 states as an input and pass this through a three-layer (1025, 256, 128) neural network that then outputs two categorical heads, one for acceleration and one for steering. The acceleration head is binned into 15 bins equally spaced between $\left[-6, 6 \right] \frac{\text{m}}{\text{s}^2}$ and steering to 43 equally spaced bins between $\left[-0.7, -0.7 \right] \text{radians}$. As there is no corresponding head tilt in the expert actions we extend the angle of the visibility cone to $\pi \, \text{radians}$. Note that this violates a rule on the cone shape set forth in Sec.~\ref{sec:rules}; figuring out how to add head tilt to imitation agents remains an open question.

\subsection{Hyperparameters}
For APPO we use almost all the default hyperparameters of SampleFactory~\cite{petrenko2020sample} at commit aed6cc92a7eb3510c4d4bcfac083ced07b5222f9. However, we use a batch size of $7168$ and scale the observations by $10.0$.

For Imitation Learning we use a learning rate of $3 \cdot 10^{-4}$, a batch size of $512$, and stack a total of $5$ states together to endow the agents with memory. 

\subsection{State space Details}
Since likely control architectures require the vector to be of a fixed size, we select a subset of visible road points and objects if there are too many and pad with a vector of $0.0$'s if there are too few. We sort both the road points, objects, and stop signs by distance and return the $500, 16, 4$ closest ones respectively where all three values are configurable by user. Using these values, the total state dimensionality is $6727$. We note that this is a fairly high dimensional state and intentionally do not prune it to be smaller as dealing with this high dimensional input is a meaningful part of the benchmark. However, since we expect that users will likely want to construct their own types of states, utility functions are also provided to allow users to query the set of observed road points and objects.
\subsection{Features}
The features of the ego object are:
\begin{itemize}[noitemsep]
    \item[--] The speed of the object.
    \item[--] The distance and angle to the goal.
    \item[--] Its width and length.
    \item[--] The relative speed and heading to the target speed and heading.
\end{itemize}

Road object features are:
\begin{itemize}[noitemsep]
    \item[--] The speed of the object.
    \item[--] The angle between the velocity vectors of the object and of the ego vehicle.
    \item[--] Its width and length.
    \item[--] The angle and distance to the road object's position relative to the ego vehicle.
    \item[--] Its heading relative to the ego vehicle.
    \item[--] A one-hot vector indicating whether the object is a vehicle, pedestrian, or cyclist.
\end{itemize}

As per the VectorNet representation, each road point is part of a discretized polyline with an approximate discretization size of $0.5$ meters. To ensure that any vehicle is able to discern which road points are connected to each other, we include a vector pointing to the neighbor of the point in the polyline. As with the road object representation, all points are in the frame of the observing vehicle.
Road point features are:
\begin{itemize}[noitemsep]
    \item[--] The angle and distance to the road point's position relative to the ego vehicle.
    \item[--] A 2D element representing the vector pointing from the current road-point to its neighbor in the polyline.
    \item[--] A one-hot vector indicating whether the road point is a lane-center, a road line, a road-edge that can be collided with, a stop-sign, a crosswalk, a speed-bump, or is of unknown type.
\end{itemize}

\section{Zero-shot coordination}
\label{sec:app_zsc}
\begin{figure}
    \centering
     \begin{subfigure}[b]{0.48\textwidth}
         \centering
         \includegraphics[width=\textwidth]{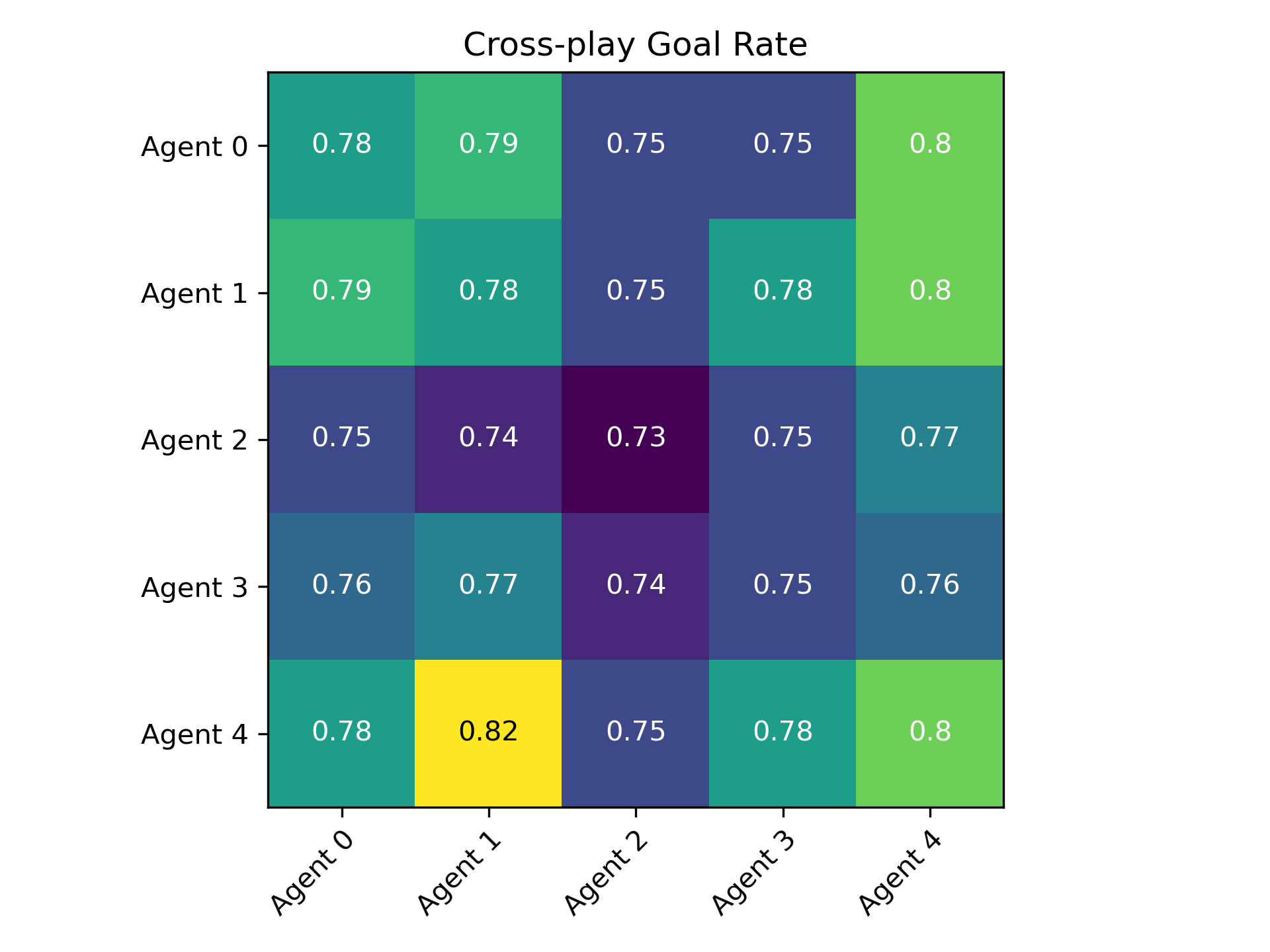}
     \end{subfigure}
          \begin{subfigure}[b]{0.48\textwidth}
         \centering
         \includegraphics[width=\textwidth]{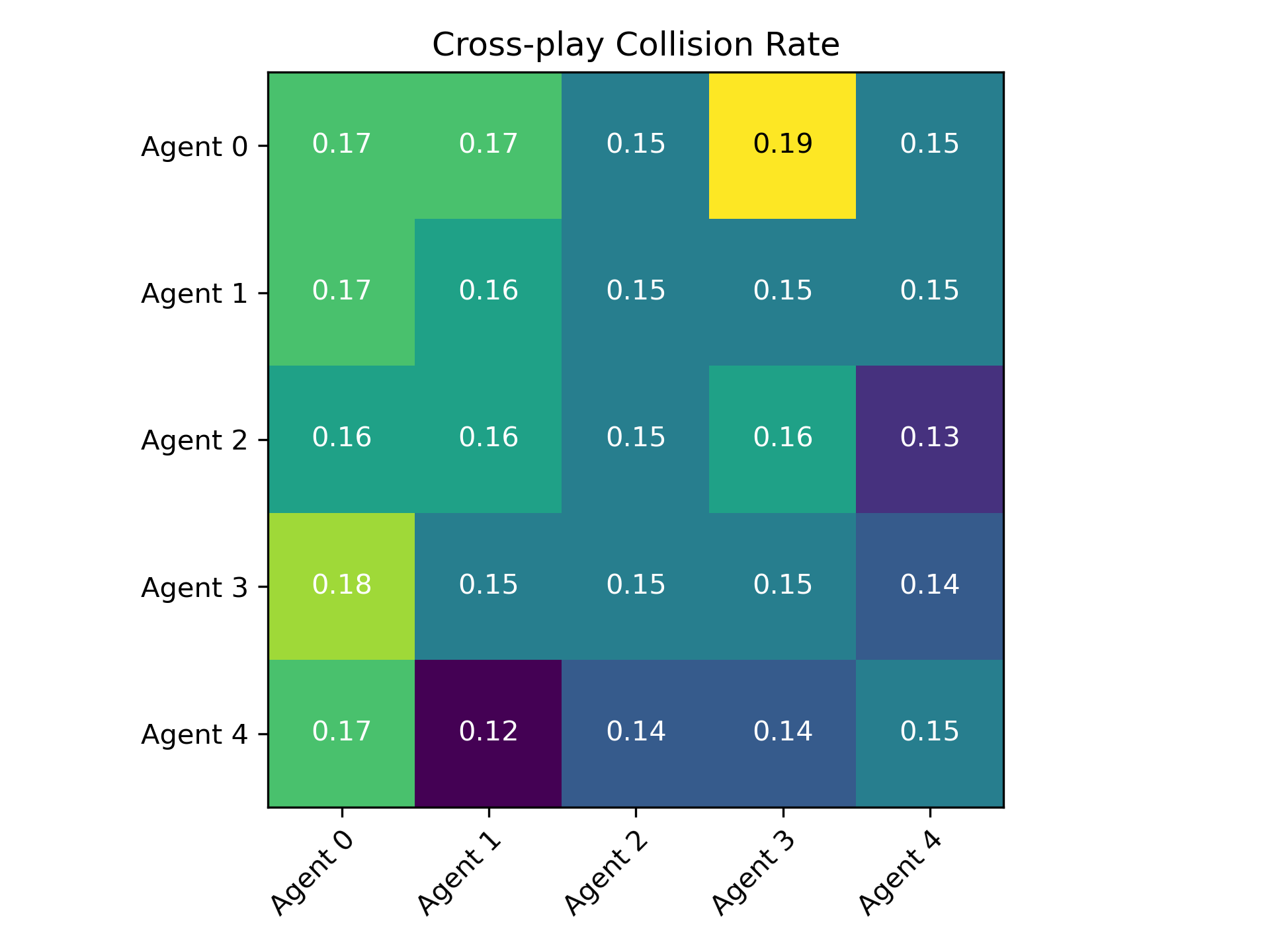}
     \end{subfigure}
    \caption{Goal rate (left) and collision rate (right) of agents when 50\% are sampled from one side and 50\% from another. The diagonal represents the baseline performance of the seed while element i, j represents seed i playing vs. seed j.}
    \label{fig:zsc}
\end{figure}

We investigate whether the agents are learning incompatible conventions across seeds (ZSC) by taking the seeds from our best performing agent on the test set (the agent trained on all files) and sampling half the agents from one seed and half from another. We then perform this procedure across all 5 seeds, running each pair on the same set of 100 files. The results of this are summarized in Fig.~\ref{fig:zsc} and appear to indicate that at the current level of agent performance there does not appear to be a zero-shot coordination issue. However, it is still possible that one might emerge as the agents become more capable and learn arbitrary symmetry-breaking conventions that are not compatible across agents. This argument lines up with the results of Sec.~\ref{sec:failure_modes} which shows that the agents fail at high rates in interactive scenes; it may be necessary to get to lower failure rates in such settings for ZSC issues to emerge.

\section{Ablation of the targets}

The experiments described in the main body of the text provide all the agents with a target position, speed, and heading for their goal. In this section we investigate the effect on some of the metrics of relaxing the speed target requirement. Based on the results from Sec.~\ref{sec:analysis_baseline}, we only run experiments using $10000$ training files as this appears to be sufficient to close the train-test gap and allows us to minimize the computational requirements of these ablations. 

Table~\ref{table:comparison_ablation} lists a comparison between the original results and the ablation on target speed. Note that since the target requirements are relaxed, the target may become easier to achieve which in turn drives up the goal rate. However, we observe the following features. First, the collision rate increases sharply relative to APPO and the goal rate decreases markedly. It is possible that the speed target serves as a hint or curriculum that makes solving some of the scenes easier; for example some scenes could be solvable by simply linearly interpolating between the agent speed and the target speed.
Since the requirements for reaching the goal are less constrained which allows for a greater range of possible trajectories to reach the target, the ADE increases and the FDE increases significantly. The sharp increase in FDE likely occurs due to the agent being able to reach its goal at higher speed and since the post-goal behavior is out-of-distribution, the agent attempts often unsuccessfully to route itself back to its goal.

\begin{table}[ht]
\caption{Overview of metrics across methods for an $8$ second rollout on the test set.} % title of Table
\centering % used for centering table
{
\begin{tabular}{c c c c c} % centered columns (4 columns)
\hline\hline %inserts double horizontal lines
Algorithm\rule{0pt}{1ex} & Collision Rate (\%) & Goal Rate (\%) & ADE (m) & FDE (m) \\  % inserts table
%heading
\hline % inserts single horizontal line
Expert Playback & $4.9$ & 100 & 0 & 0 \\ % inserting body of the table
APPO & $20.3 \pm 0.8$ & $71.7 \pm 0.7$ & $3.1 \pm 0.2$ & $6.1 \pm 0.3$ \\
APPO w/o target speed & $29.3 \pm 1.2$ & $64.9 \pm 0.9$ & $4.3 \pm 0.3$ & $12.6 \pm 0.9$ \\
BC & $38.2 \pm .1$ & $25.3 \pm 0.1$ & $5.6 \pm 0.1$ & $9.2 \pm 0.1$\\
\hline %inserts single line
\end{tabular}}
\label{table:comparison_ablation} % is used to refer this table in the text
\end{table}

\color{black}
\section{Vehicle Model}
\label{sec:vehicle_model}
Our vehicles are driven by a kinematic bicycle model~\cite{rajamani2011vehicle} which uses the center of gravity as reference point. The dynamics are as follows. Here $(x_{t}, y_{t})$ stands for the coordinate of the vehicle's position at time $t$, $\theta_{t}$ stands for the vehicle's heading at time $t$, $v_{t}$ stands for the vehicle's speed at time $t$, $a$ stands for the vehicle's acceleration and $\delta$ stands for the vehicle's steering angle. $L$ is the distance from the front axle to the rear axle (in this case, just the length of the car) and $l_{r}$ is the distance from the center of gravity to the rear axle. Here we assume $l_{r} = 0.5L$.

\begin{align*}
    \Dot{v} &= a \\
    \Bar{v} &= \text{clip}(v_t + 0.5 \ \Dot{v} \ \Delta t, -v_\text{max}, v_\text{max}) \\
    \beta &= \tan^{-1}\left(\frac{l_{r} \ \tan(\delta)}{L}\right) \\
          &= \tan^{-1}(0.5 \ \tan(\delta)) \\
    \Dot{x} &= \Bar{v} \ cos(\theta + \beta) \\
    \Dot{y} &= \Bar{v} \ sin(\theta + \beta) \\
    \Dot{\theta} &= \frac{\Bar{v} \ \cos(\beta) \ \tan(\delta)}{L}
\end{align*}

We then step the dynamics as follows:

\begin{align*}
    x_{t+1} &= x_{t} + \Dot{x} \ \Delta t \\
    y_{t+1} &= y_{t} + \Dot{y} \ \Delta t \\
    \theta_{t+1} &= \theta_{t} + \Dot{\theta} \ \Delta t \\
    v_{t+1} &= \text{clip}(v_t + \Dot{v} \ \Delta t, -v_\text{max}, v_\text{max})
\end{align*}

\label{app:veh_model}

\section{Calculating Infeasible Goal Statistics}
\label{sec:infeasible}
For computing the percentage of vehicles that must cross a road edge to get to their goal, we use the following procedure. We take the vehicles and shrink their width by 0.1 and their length by 0.3. These vehicles are very thin and so do not accidentally collide with a road edge due to small errors in their position. The scaling of the length ensures that the vehicles are not placed in a starting position where they have collided. Using this procedure we calculate the $3\%$ statistic.

\section{Computing the simulator speed}
\label{app:sec:speed}
\begin{figure}
    \centering
     \begin{subfigure}[b]{0.48\textwidth}
         \centering
         \includegraphics[width=\textwidth]{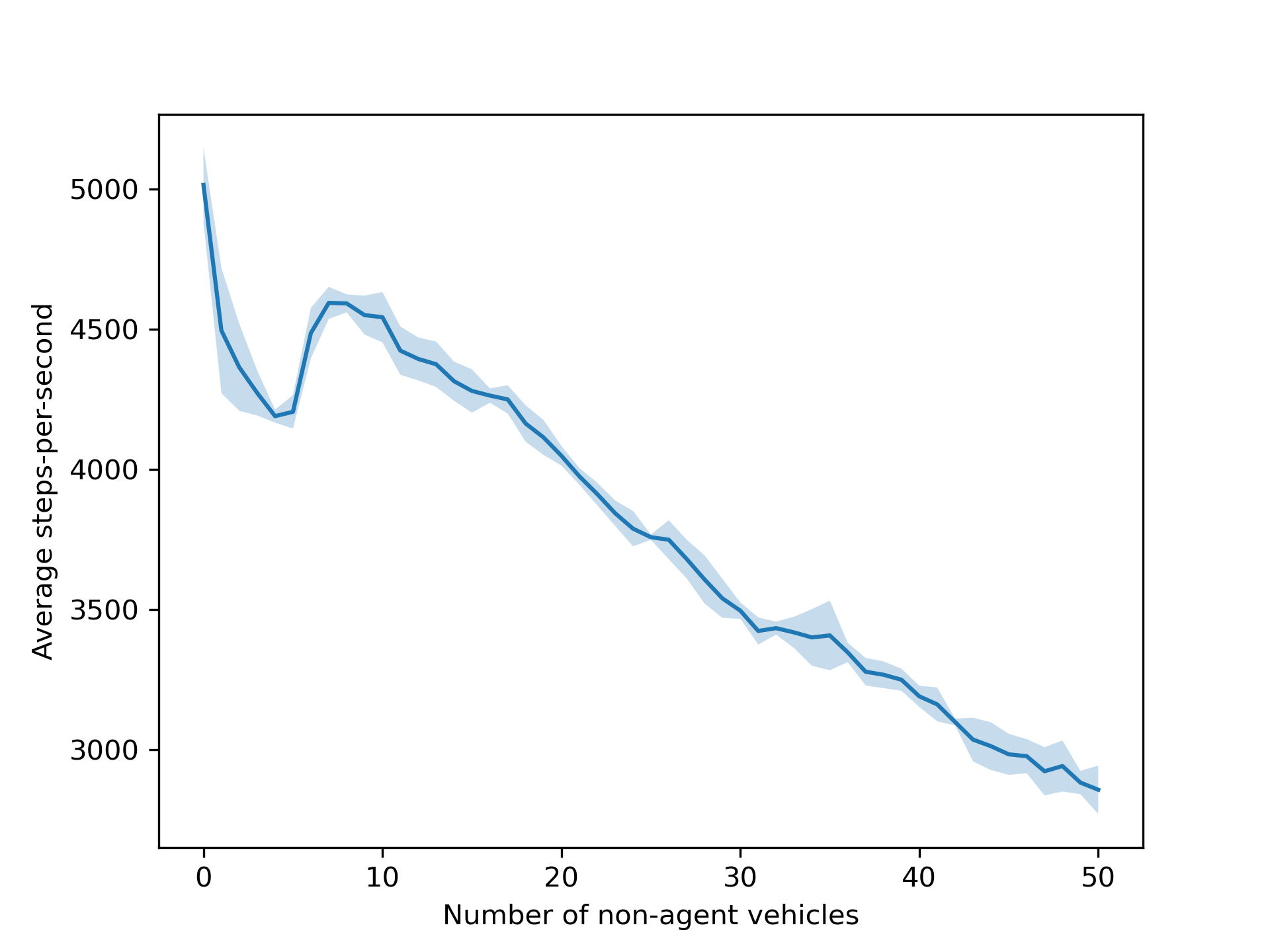}
     \end{subfigure}
          \begin{subfigure}[b]{0.48\textwidth}
         \centering
         \includegraphics[width=\textwidth]{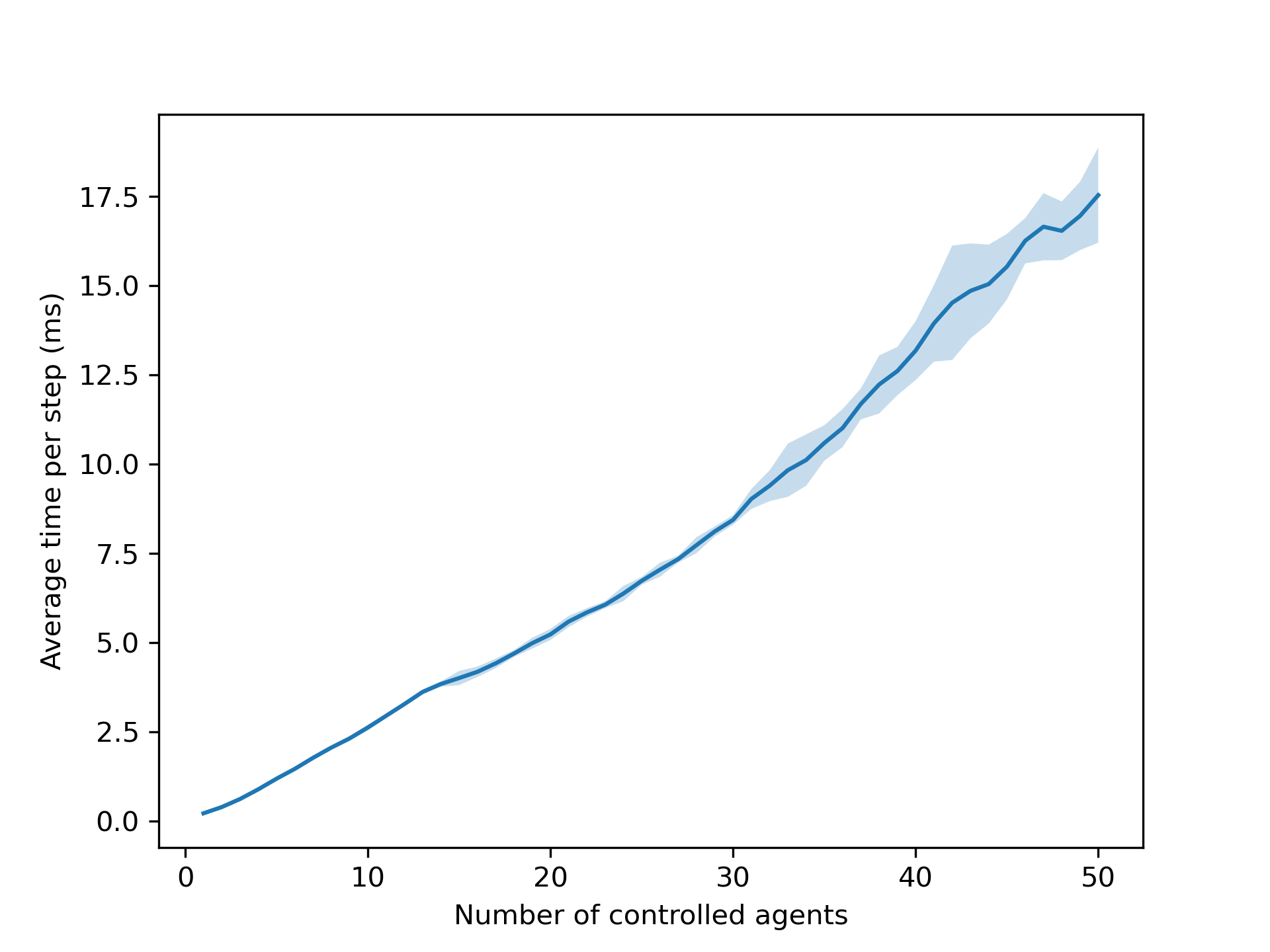}
     \end{subfigure}
    \caption{(Left) \textbf{Single-agent}: average steps-per-second as a function of the number of non-agent vehicles in the scene, where each step corresponds to computing observations for a single agent and stepping the simulation. (Right) \textbf{Multi-agent}: average time to compute observations and step all the controlled agents in a scene as a function of the number of agents in the scene. Shading corresponds to the standard deviation of the mean. The x-axis is cut at 50 as there are few scenes above this value.}
    \label{fig:fps}
\end{figure}

We use two procedures for determining the overall speed of our simulator. The first measures the time to compute observations and step a single agent, which is what is reported as the simulator speed in the paper, and the second presents the time to compute observations and step a group of agents. In the first procedure, we randomly select a set of 2000 files and step through them from the first to the last time-step. At each time-step, we sample a random agent, set the remaining agents to be replayed via expert data, compute the observation for the sampled agent, compute a random action for the sampled agent, and then step the simulator forwards. We repeat this procedure 5 times and report average results and standard deviation of the mean values. Over this dataset and procedure, the single-agent steps-per-second (SPS) is around $3400$. This is a slightly modified version of the procedure used to compute FPS in MetaDrive~\cite{li2021metadrive} to account for variations in the number of agents available in a scene as well as the lack of rule-based agents in our simulator. With 10 non-agent vehicles we achieve a SPS of $4543$, $4049$ at 20, and $3497$ at 30. The left side of Fig.~\ref{fig:fps} shows the scaling with the number of other vehicles in the scene which includes the costs of computing partial observability, collision checking, and stepping the bicycle model.

We also perform a multi-agent variant of this procedure to test the scaling of our simulator with the number of controlled agents. Here we sample observations for all potential controlled agents in a scene and then take a simulator step.  Again, we repeat this procedure 5 times, each time sweeping across all scenes in the $2000$ files as defined in the prior procedure, and report average results and standard deviation of the mean values over the runs. The right side of Fig.~\ref{fig:fps} shows the scaling of the time to perform this procedure as a function of the number of controlled agents in the scene. We observe a seemingly linear scaling with agent number. There are about $8.5$ agents on average and we achieve an average per-agent SPS of $452$, allowing us to construct about $3800$ frames of experience per-second.

%In line with this, at eight vehicles (the average number of controlled agents in the system) we observe an FPS of approximately $500$ for each agent, allowing us to construct $4000$ frames of experience per-second for the $8$-agent system. On average, swept across all scenes in the $1000 files$, we achieve a per-agent FPS of $480$.

\section{License Details and Accessibility}
\label{sec:license}
Our code is released under an MIT License. The Waymo Motion dataset is released under a Apache License 2.0. The code is available at \url{https://github.com/facebookresearch/nocturne}.

\end{document}